\documentclass[aps,prd,superscriptaddress,showkeys,12pt]{revtex4-2}
\usepackage{inputenc}
\usepackage{amsmath,amsfonts,amssymb}
\usepackage{graphicx, color}
\usepackage{natbib}
\usepackage{hyperref}
\usepackage{setspace,comment}

\usepackage[dvipsnames]{xcolor} 

\usepackage{float}
\usepackage{orcidlink}
\usepackage{booktabs} 
\newcommand{\be}{\begin{equation}}
\newcommand{\ee}{\end{equation}}
\newcommand{\bea}{\begin{eqnarray}}
\newcommand{\eea}{\end{eqnarray}}

\begin{document}


\title{Late-time acceleration without a vacuum term in $f(R,L_m)$ gravity: scaling deSitter dynamics and parameter constraints}

\author{Luciano Navarro-Coydán\orcidlink{0000-0002-0000-0000}}
\affiliation{Departamento de Física Teórica, Facultad de Física, Universidad de Valencia, 46100 Burjassot, Valencia, Spain}

\author{J. Alberto Vázquez \orcidlink{0000-0002-7401-0864}}
\affiliation{Instituto de Ciencias Físicas, Universidad Nacional Autónoma de México, 62210, Cuernavaca, Morelos, México}

\author{Israel Quiros\orcidlink{0000-0002-0120-0624}}
\affiliation{Dpto. Ingeniería Civil, División de Ingeniería, Universidad de Guanajuato, 36000, Gto., México.}

\author{Ricardo García-Salcedo\orcidlink{0000-0003-0173-5466}} \thanks{Corresponding author: rigarcias@ipn.mx}
\affiliation{Centro de Investigación en Ciencia Aplicada y Tecnología Avanzada, Unidad Legaria del Instituto Politécnico Nacional, Legaria 694, 11500, Ciudad de México, México.}
\affiliation{Universidad Internacional de Valencia - VIU, 46002, Valencia, Spain.}

\date{\today}

\begin{abstract}
We investigate late-time cosmic acceleration in $f(R,L_m)$ gravity driven by nonlinear matter contributions, focusing on the class
$f(R,L_m)=R/2+c_1 L_m+c_n L_m^{n}+c_0$ with the explicit choice $L_m=\rho_m$ and an uncoupled radiation sector. We analyze two realizations: (i) Case A: $f(R,L_m)=R/2+\beta \rho_m^{n}+\gamma$, where $\gamma$ acts as a vacuum term, and (ii) Case B: $f(R,L_m)=R/2+\beta \rho_m+\gamma \rho_m^{n}$, where the nonlinear sector can mimic dark energy without an explicit cosmological constant. For each case, we construct a bounded autonomous system, classify all critical points and their stability, and compute cosmographic diagnostics. The phase-space analysis shows that Case A reproduces the standard radiation$\to$matter$\to$de~Sitter sequence only for $n\gtrsim 4/5$, with acceleration essentially enforced by the vacuum term.
In contrast, Case~B admits a qualitatively distinct and phenomenologically appealing branch: for $0<n<1/2$ the system possesses a physical \emph{scaling} de~Sitter future attractor inside the bounded simplex, yielding radiation$\to$matter$\to$acceleration with $q=-1$ and $\omega_{\rm eff}=-1$ and without introducing $c_0$. We confront both models with background data (CC, Union3, DESI BAO, plus a BBN prior on $\Omega_b h^2$) using nested sampling and perform model comparison via Bayesian evidence and AIC/BIC. The full data combination constrains $n=1.08\pm0.05$ in Case A and $n=0.05\pm0.10$ in Case B (68\% CL), the latter lying within the accelerating window while remaining statistically consistent with $\Lambda$CDM kinematics at the background level. We also record minimal consistency conditions for stability (tensor no-ghost and luminal propagation) and motivate a dedicated perturbation-level analysis as the next step to test growth and lensing observables.
\end{abstract}

\keywords{cosmology, dark energy, dynamical systems, modified gravity, $f(R, L_m)$ gravity}

\maketitle

\section{Introduction}\label{sec-intro}  

For the last 25 years, extensive research has strongly suggested that our universe is undergoing an accelerating expansion \cite{riess1998observational, perlmutter1998discovery, perlmutter2003supernovae, riess2004type, percival2010baryon, spergel2003first, koivisto2006dark, daniel2008large, bennett2003microwave}. The concordance model recognizes two phases of accelerated expansion: the primordial phase at the beginning of the universe and the phase we are currently experimenting. To explain this recent acceleration within the framework of General Relativity (GR), an antigravitational force, hypothesized to be dark energy (DE), is proposed. The most common description of DE is the cosmological constant $\Lambda$ (also referred to as vacuum energy), which is affected by the well-known cosmological constant problem (CCP) \cite{lopez2018problems}. Another component of the concordance model, exotic dark matter (DM), was introduced to explain phenomena such as flattened galactic rotation curves \cite{begeman1991extended, borriello2001dark} and the formation of large-scale structures \cite{hoekstra2002current}. Both components are consistent with observational data, but it is well known that, although the $\Lambda$CDM paradigm can explain many observations with high precision, it suffers from several well-known drawbacks \cite{martin2012everything, choi4problems}.  

An alternative way to explain the accelerated expansion is to generalize the Einstein-Hilbert action of GR. Various theories have emerged attempting to explain both DM and DE as unified phenomena and to address other drawbacks of the $\Lambda$CDM theory. This has led to many ``Extended Gravity Theories" that generalize the GR action. Notable examples include Lovelock theories \cite{lovelock1971einstein}, Gauss-Bonnet theories \cite{cognola2006dark, de2012stability}, scalar-tensor theories \cite{brans1962mach, cembranos2009quantum, quiros-rev}, Tensor-Vector-Scalar (TeVeS) gravity \cite{ford1989inflation}, and theories with extra dimensions \cite{giudice1999quantum}, among others.  

A prominent example is $f(R)$ gravity, which replaces the Einstein-Hilbert Lagrangian by a general function $f(R)$ of the Ricci scalar. A key feature of this theory is its ability to describe early- and late-time accelerated expansion without requiring a cosmological constant \cite{saridakis2021modified}. Recent analyzes show that $f(R)$ models are compatible with Planck 2018 constraints on late-time evolution and avoid unwanted DE oscillations \cite{oikonomou2021unifying}. These models also comply with solar system tests \cite{faraoni2006solar, zhang2007behavior}, and some extensions involve non-minimal or arbitrary coupling between curvature and matter \cite{bertolami2008nonminimal, harko2008modified}.  

A broader extension of $f(R)$ gravity is the so-called $f(R, L_m)$ theory, where the gravitational action includes an arbitrary function of both the Ricci scalar $R$ and the matter Lagrangian $L_m$. This idea, originally proposed in \cite{harko2010f, harko2014generalized}, generalizes previous dark coupling models of the form $S = \int \left[ f_1(R) + (1 + \lambda f_2(R)) L_m \right] \sqrt{-g} d^4x$ \cite{bertolami2007extra}. In $f(R, L_m)$ gravity, the action takes the form $S = \int f(R, L_m) \sqrt{-g} d^4x$, and the dynamics is influenced by direct matter-curvature interactions, which makes the theory particularly relevant for cosmology \cite{harko2010matter,harko2012geodesic,wang2012energy}.  

In vacuum ($T_{\mu\nu} = 0$), $f(R, L_m)$ reduces to standard $f(R)$ gravity. However, in the presence of matter, it predicts the existence of an extra force orthogonal to the four-velocity of matter, resulting in non-geodesic motion \cite{harko2015gravitational}. This departure from GR has significant implications for cosmic evolution and conservation laws. Moreover, the flexibility of $f(R, L_m)$ allows it to accommodate a wide range of equations of state, making it a promising framework to address open questions such as the nature of dark energy and dark matter \cite{nesseris2009newton, faraoni2007viability, faraoni2009lagrangian, harko2015gravitational}.  

Recent literature in $f(R,L_m)$ cosmology has been predominantly background-focused—typically combining Supernovae Type~Ia (SNe Ia), Baryon Acoustic Oscillations (BAO) and Cosmic Chronometers (CC, measuring $H(z)$)—with limited use of dynamical-systems/attractor methods and virtually no linear-perturbation confrontations (growth via RSD/$f\sigma_8$, nor lensing diagnostics $\mu(a,k),\Sigma(a,k),\eta$). In several cases, the choice of matter Lagrangian (e.g., $L_m=\rho$ vs.\ $L_m=-\rho$) is either implicit or underspecified, which complicates comparisons across models and datasets. These gaps motivate a careful separation between background-level viability and full perturbative consistency, and they clarify the value of a clean, bounded phase-space analysis at the background level before advancing to structure-growth tests.  

An important choice in these models is the form of the matter Lagrangian $L_m$. In many cosmological applications, it is taken to be $L_m = \rho_m$, which implies a non-conservation of the energy-momentum tensor due to the matter-curvature coupling. This choice highlights the direct role of the energy density in the interaction, and it is suitable for comparison with observational constraints \cite{harko2010f, harko2014generalized, jaybhaye2022cosmology, garg2024cosmological}. Other studies have also considered $L_m = -\rho_m$ to preserve covariant conservation \cite{faraoni2009lagrangian, bertolami2008nonminimal}, but we adopt the first choice here, following the recent literature and its physical motivation. We explicitly track this choice throughout to avoid ambiguities when comparing with background and perturbation-level observables.  

Recent investigations in the context of $f(R, L_m)$ gravity have introduced power-law matter terms of the form $\rho_m^n$ to mimic dark energy dynamics, without invoking a cosmological constant. Models such as $f(R,L_m) = \frac{R}{2} + \rho_m^n + \gamma$ and $f(R,L_m) = \frac{R}{2} + \rho_m^n$ have been studied in \cite{jaybhaye2022cosmology, shukla2023dynamical}, showing consistency with observational data, including Hubble expansion rate and supernova compilations \cite{garg2024cosmological, patil2023flrw}. These studies suggest that nonlinear energy density terms can effectively drive late-time cosmic acceleration and transition behaviors.  

A phenomenologically related class are the \emph{Cardassian models} of Freese and Lewis \cite{freese2002cardassian}, in which the Friedmann equation is modified as $H^2=\tfrac{8\pi G}{3}\rho_m+B\rho_m^{n}$. For $0<n<1$, the extra term can dominate at low densities and trigger late–time acceleration without a cosmological constant. Analyzes with the SNe~Ia, BAO and Cosmic Microwave Background (CMB) datasets have shown that the values $n=\mathcal{O}(0.2\text{--}0.3)$ can reproduce the deceleration–acceleration transition around $z\sim 0.5\text{--}0.7$ \cite{zhu2003constraints,sen2003observational}. However, the Cardassian term is introduced at the level of the background as an \emph{ad hoc} modification, without a variational underpinning. In contrast, within the $f(R,L_m)$ framework, power–law matter contributions $\propto \rho_m^{n}$ arise naturally from non–minimal curvature-matter couplings in the action \cite{harko2010f}, allowing a consistent treatment of background expansion, dynamical stability, and modified conservation laws. This connection effectively roots Cardassian–type phenomenology in a modified–gravity setting and motivates the models studied here.  

To organize dimensions and parameter identifications across different realizations, it is convenient to introduce a unified template
\begin{equation}
    f(R,L_m) \;=\; \alpha\,\frac{R}{2} \;+\; c_1 L_m \;+\; c_n L_m^{n} \;+\; c_0,
\end{equation}
with $L_m=\rho_m$. Throughout this work, we adopt units $8\pi G=c=1$, so that the Ricci scalar and the matter Lagrangian have the same physical dimension, $[R]=[\rho_m]\sim H^2$. In this normalization, $\alpha$ and $c_1$ are dimensionless coefficients that multiply the linear terms in $R$ and $L_m$, respectively. The constant term $c_0$ carries the same mass dimension as $R$ (that of a cosmological constant), while the coefficient of the nonlinear contribution must compensate for the power of $L_m$, i.e.
\begin{equation}
    [c_n] = [R]^{\,1-n},
\end{equation}
so that $c_n L_m^n$ has the same overall dimension as $R$. In practice, one can factor out the dimensional part of $c_n$ into a reference density scale (or, equivalently, into a rescaling of $\rho_m$), and work with effective dimensionless couplings in the background and dynamical analyzes.

The various models considered in the literature and in this work correspond to particular choices of $(c_1,c_n,c_0)$. For instance, the power-law models $f(R,L_m)=\tfrac{R}{2}+\rho_m^{n}+\gamma$ studied in \cite{jaybhaye2022cosmology, shukla2023dynamical} correspond to $(c_1=0,\,c_n=1,\,c_0=\gamma)$ with $L_m=\rho_m$, where the dimensional factor of $c_n$ is absorbed into the normalization of $\rho_m$. In our analysis, \emph{Case A} and \emph{Case B} are understood as two specific realizations of the above template. For Case A, we take $L_m=\rho_m$ and
\begin{equation}
    c_1^{(A)} = 0, \qquad c_n^{(A)} = \beta, \qquad c_0^{(A)} = \gamma,
\end{equation}
so that
\begin{equation}
    f(R,L_m)=\tfrac{R}{2}+\beta\rho_m^{n}+\gamma. \label{fCA}
\end{equation}
For Case B, we consider
\begin{equation}
    c_1^{(B)} = \gamma, \qquad c_n^{(B)} = \beta, \qquad c_0^{(B)} = 0,
\end{equation}
which leads to
\begin{equation}
    f(R,L_m)=\tfrac{R}{2}+\gamma\rho_m+\beta\rho_m^{n}. \label{fCB}
\end{equation}

In both cases, the couplings $(\beta,\gamma)$ should be understood as effective parameters defined after fixing a reference density scale. 
In our units $8\pi G=c=1$, one has $[R]=[\rho_m]\sim H^2$. Hence, in Case A, the constant term $\gamma$ carries dimensions of $H^2$ (it is vacuum-energy--like), whereas in Case B, the coefficient of the linear matter term is dimensionless. For the background dynamics and the observational analysis, it is therefore convenient to trade the original couplings for dimensionless density amplitudes (e.g.\ $\Omega_{\gamma,0}$ in Case A and $\Omega_{NL,0}$ in Case B), which are the quantities directly entering the normalized Friedmann equation and the likelihood. Accordingly, at the background level, the physically relevant information is encoded; this implies that the physically relevant information is encoded in the exponent $n$ and in the (dimensionless) density amplitudes appearing in the Friedmann equation, while different choices of $(c_1,c_n,c_0)$ related by overall rescalings of $\rho_m$ lead to equivalent autonomous systems. This unified parameterization will be useful later when discussing identifiability and reparameterizations in terms of dimensionless density amplitudes at the background level.

In this work, we explore two specific models within the $f(R, L_m)$ framework. The first model considers a matter Lagrangian of the form given in Eq. ({\ref{fCA}}), $f(R,L_{m}) = \frac{R}{2} + \beta \rho_m^n + \gamma$, 
where $\gamma$ plays the role of a cosmological constant and $\rho_m^n$ acts as a nonlinear coupling term. This model is not intended to explain the late-time accelerated cosmic dynamics, since it contains a cosmological constant term, but it allows us to explore how the combination of a vacuum energy term and a nonlinear matter contribution can influence cosmic dynamics. Additionally, a term is included that encodes Maxwell's electromagnetic radiation. 

The second model contains a linear matter term and retains the nonlinear contribution, as shown in Eq. (\ref{fCB}), $f(R,L_{m}) = \frac{R}{2} + \gamma \rho_m + \beta \rho_m^n$, 
where Maxwell's electromagnetic radiation is also explicitly included through $\rho_r$. Here, the term $\rho_m^n$ is not interpreted as a fluid with an exotic equation of state but rather as an effective dark energy component arising from nonlinear interactions of matter. In other words, the nonlinear term can generate late-time acceleration without resorting to a cosmological constant. In this sense, our analysis complements recent observational reconstructions of $f(R,L_m)$ models, such as those obtained by Devi \emph{et al.} \cite{devi2025late}, who employed Gaussian Processes to infer viable forms of $f(L_m)$ directly from $H(z)$ data. Their results indicate that both power-law ($f(L_m)\propto L_m^{b_1}$ with $b_1\simeq 0.02$) and exponential parameterizations are consistent with current cosmological observations and admit late-time de Sitter attractors. Thus, while their approach provides empirical evidence for the viability of nonlinear $L_m$ couplings, our work offers a dynamical justification of how such terms can drive the standard radiation--matter--acceleration sequence, linking phenomenological reconstructions with a theoretical foundation in modified gravity.

Our approach involves a combined study comprising a dynamical-systems analysis and subsequent parameter estimation. The dynamical-systems analysis is designed to identify and characterize the phase-space critical points. This enables us to examine the asymptotic behavior of solutions and understand how the nonlinear term $\rho_m^n$ affects cosmic dynamics. Particular attention is given to the possibility of reproducing $\Lambda$CDM-like dynamics or alternative acceleration mechanisms. Meanwhile, the (background-only) parameter estimation allows us to delineate observationally allowed regions for the free parameters $(n,\beta,\gamma)$, reserving structure-growth and lensing tests for a dedicated perturbative analysis.

The novelty of the present study lies in a systematic comparison of two classes of $f(R, L_m)$ models: one with an explicit vacuum energy term and the other with a purely dynamical effective dark energy component arising from nonlinear effects of matter. In both cases, we constrain the parameter $n$ using insights from phase space analysis and connect the results to observationally motivated ranges. Given the current literature’s emphasis on background fits and the scarcity of perturbative confrontations, we clearly separate the scopes: the present paper establishes background viability (bounded phase space, cosmography, $c_a^2$ at critical points, and background constraints). In contrast, a companion future work will present the complete linear-perturbation treatment (rest-frame $c_s^2$, no-ghost/no-gradient conditions, and effective functions $\mu(a,k),\Sigma(a,k),\eta$) jointly constrained with RSD/$f\sigma_8$ and lensing data.

\section{The $f(R, L_m)$ gravity model}\label{sect-setup}

In the $f(R, L_m)$ gravity framework, the modified action is given by \citep{harko2010f}:

\begin{equation} S = \int \sqrt{-g} \, d^4x \, f(R, L_m),\end{equation} where $R$ is the Ricci scalar, $L_m$ is the matter Lagrangian density, and $g$ is the determinant of the metric tensor $g_{\mu\nu}$. We adopt the convention $8\pi G = c = 1$, where $G$ and $c$ denote the Newtonian gravitational constant and the speed of light, respectively.

In this work, we consider an additional uncoupled Maxwell radiation sector represented by a term $L_r$, leading to the total action:
\begin{equation}
    S = \int \sqrt{-g} \, d^4x \left[ f(R, L_m) + L_r \right]. \label{Act2}
\end{equation}

The corresponding energy-momentum tensors for matter and radiation are defined as:
\begin{align}
    \mathcal{T}^{(m)}_{\mu\nu} &= -\frac{2}{\sqrt{-g}} \frac{\delta(\sqrt{-g} L_m)}{\delta g^{\mu\nu}}, \label{EMT} \\
    \mathcal{T}^{(r)}_{\mu\nu} &= -\frac{2}{\sqrt{-g}} \frac{\delta(\sqrt{-g} L_r)}{\delta g^{\mu\nu}}. \label{ERT}
\end{align}

Varying the action \eqref{Act2} with respect to the metric tensor $g_{\mu\nu}$ yields the field equations for the $f(R, L_m) + L_r$ theory:
\begin{equation}\label{FE}
    f_R R_{\mu\nu} + \left( g_{\mu\nu} \Box - \nabla_\mu \nabla_\nu \right) f_R - \frac{1}{2} \left( f - f_{L_m} L_m \right) g_{\mu\nu} = \frac{1}{2} f_{L_m} \mathcal{T}^{(m)}_{\mu\nu} + \frac{1}{2} \mathcal{T}^{(r)}_{\mu\nu},
\end{equation}
where $f_R \equiv \partial f / \partial R$ and $f_{L_m} \equiv \partial f / \partial L_m$.

Taking the covariant divergence of Eq. \eqref{FE}, and using the Bianchi identity, we obtain the following conservation equations:
\begin{equation}
    \nabla^\mu \mathcal{T}^{(m)}_{\mu\nu} = 2 \nabla^\mu \ln(f_{L_m}) \frac{\partial L_m}{\partial g^{\mu\nu}},  \quad \quad \quad
    \nabla^\mu \mathcal{T}^{(r)}_{\mu\nu} = 0. \label{CDM}
\end{equation}

We now assume a homogeneous and isotropic universe described by the spatially flat FLRW metric:
\begin{equation}\label{FLRW}
    ds^2 = -dt^2 + a^2(t)(dx^2 + dy^2 + dz^2),
\end{equation}
where $a(t)$ is the cosmic scale factor. For a perfect fluid, we choose $L_m = \rho_m$ for matter and $L_r = \rho_r$ for radiation \citep{harko2015gravitational, faraoni2009lagrangian}. The corresponding energy-momentum tensors take the form:
\begin{align}
    \mathcal{T}^{(m)}_{\mu\nu} &= (\rho_m + p_m) u_\mu u_\nu + p_m g_{\mu\nu}, \label{EMT1} \\
    \mathcal{T}^{(r)}_{\mu\nu} &= (\rho_r + p_r) u_\mu u_\nu + p_r g_{\mu\nu}, \label{EMT2}
\end{align}
where $u^\mu = (1, 0, 0, 0)$ is the four-velocity of the comoving observer, satisfying $u^\mu u_\mu = -1$. The Ricci scalar is given by
\begin{equation}
    R = 6(\dot{H} + 2H^2),
\end{equation}
where $H = \dot{a}/a$ is the Hubble parameter.

Substituting into the field equations, the modified Friedmann equations for this theory are obtained as \citep{jaybhaye2022cosmology, koussour2024bouncing}:
\begin{align}
    3H^2 f_R + \frac{1}{2} \left(f - f_R R - f_{L_m} L_m\right) + 3H \dot{f_R} &= \frac{1}{2} f_{L_m} \rho_m + \frac{1}{2} \rho_r, \label{EF1} \\
    \dot{H} f_R + 3H^2 f_R - \ddot{f_R} - 3H \dot{f_R} + \frac{1}{2} \left(f_{L_m} L_m - f\right) &= \frac{1}{2} f_{L_m} p_m + \frac{1}{2} p_r, \label{EF2}
\end{align}
where dots denote derivatives with respect to cosmic time $t$.

These equations will serve as the starting point for our analysis of specific functional forms of $f(R, L_m)$, aimed at explaining late-time cosmic acceleration through nonlinear energy-density terms. To characterize the cosmological behavior of these models, we introduce a set of standard cosmological parameters.

The effective equation of state parameter, $\omega_{\text{eff}}$, and the deceleration parameter, $q$, describing the expansion dynamics are given by:
\begin{equation}
\omega_{\text{eff}} = -1 - \frac{2\dot{H}}{3H^2}, \qquad q = -1 - \frac{\dot{H}}{H^2}. \label{eq:omega_q}
\end{equation}

To further distinguish our models from the standard $\Lambda$CDM cosmology, we also consider the statefinder parameters $\{r, s\}$:
\begin{equation}
r = 1 + \frac{\ddot{H}}{H^3} + 3 \frac{\dot{H}}{H^2}, \qquad s = \frac{r - 1}{3(q - \tfrac{1}{2})}. \label{eq:rs}
\end{equation}

These parameters will be used throughout the dynamical analysis to classify critical points and interpret their physical significance in terms of deceleration, acceleration, and compatibility with observational benchmarks such as the $\Lambda$CDM model.

To assess the background response of the effective mixture induced by $f(R,L_m)$, we compute the adiabatic squared sound speed
\begin{equation}
    c_a^2 \;\equiv\; \frac{\dot p_{\rm eff}}{\dot \rho_{\rm eff}},
\end{equation}
where $\rho_{\rm eff}\equiv 3H^2$ and $p_{\rm eff}\equiv -(2\dot H+3H^2)$ follow from Eqs.~\eqref{EF1}–\eqref{EF2} after identifying the effective fluid. Operationally, $c_a^2$ measures the trajectory derivative of $\omega_{\rm eff}(a)=p_{\rm eff}/\rho_{\rm eff}$ along the background solution; it depends only on background quantities and should not be confused with the rest–frame sound speed that governs linear perturbations. 

We define \(A\equiv \dot H/H^2\), so the adiabatic background sound speed becomes
\begin{equation}
    c_{a}^{2} \;=\; -1 - \frac{2}{3}A - \frac{A'}{3A}
    \;=\; \omega_{\rm eff} - \frac{1}{3}\frac{\omega'_{\rm eff}}{1+\omega_{\rm eff}},
\end{equation}
where primes denote \(d/d\ln a\). The derivation follows from
\(\rho_{\rm eff}=3H^2\), \(p_{\rm eff}=-(2\dot H+3H^2)\) and the effective
continuity equation.

\section{Qualitative Analysis of Selected \texorpdfstring{$f(R, L_m)$}{f(R, Lm)} Cosmological Models} 

\subsection{Case A: Nonlinear Matter Coupling with Vacuum Term} \label{SecIIIA}

As a first case, we consider Eq. (\ref{fCA}):
\begin{equation}
    f(R, L_m) \;=\; \frac{R}{2} \,+\, \beta L_m^n \,+\, \gamma, \label{Polyf}
\end{equation}
where $\beta$, $\gamma$ and $n$ are free parameters. In this model, $\gamma$ acts as a vacuum term (cosmological constant–like), while the nonlinear dependence $L_m^n$ can dynamically reproduce dark–energy effects. The constant $\beta$ is introduced to cure the dimensional mismatch of the nonlinear matter term, so that $\beta L_m^n$ has the same dimensions as the curvature contribution $R$ in the units $8\pi G=c=1$. In particular, for $n\neq 1$ one may view $\beta$ as carrying units of $\rho_m^{1-n}$, allowing us to explore the impact of the nonlinear coupling strength independently of the index $n$. We also include a radiation sector, $\rho_r$, to capture early–time relativistic dynamics.

Following the usual practice in cosmology within $f(R,L_m)$, we adopt $L_m=\rho_m$, which (i) makes the direct curvature-matter interaction manifest through $f_{L_m}$ (and induces the standard non-covariant conservation of the energy-momentum tensor) and (ii) allows transparent confrontation of the model with background observables \cite{harko2010f,faraoni2009lagrangian,harko2015gravitational}. In contrast, the choice $L_m=-\rho_m$ preserves covariant conservation but will not be considered here.

Using Eqs.~(\ref{EF1})–(\ref{EF2}) with $L_m=\rho_m$ and a radiation fluid with $\omega_r=1/3$, one obtains
\bea
    3H^{2} + \gamma &=& (2n-1)\,\beta \,\rho_m^{n} + \rho_{r}, \label{1FE}\\
    2\dot{H} + 3H^{2} + \gamma &=& \beta\,\rho_m^{n} (n-1) - \beta\,n \rho_m^{n-1}p_m - \tfrac{1}{3}\rho_{r}. \label{2FE}
\eea
The (modified) continuity relations read
\bea \label{continuity_case_A}
    (2n-1)\dot{\rho}_m + 3H (\rho_m + p_m) &=& 0, \label{CM1}\\
    \dot{\rho}_r + 4H \rho_r &=& 0. \label{CR1}
\eea Hence, with $L_m=\rho_m$ the matter sector retains the perfect–fluid form $T_{\mu\nu}=(\rho_m+p_m)u_\mu u_\nu+p_m g_{\mu\nu}$. For a barotropic fluid with EOS $p_m=(\gamma_m-1)\rho_m$ ($\gamma_m$ is the barotropic index of the fluid), straightforward integration of \eqref{CM1} leads to $\rho_m\propto a^{-3\gamma_m/(2n-1)}$. Note that for $n<1/2$ the energy density of the fluid increases with cosmic expansion. GR is recovered at $n=1$. For $\gamma=\Lambda$, $p_m=0$ ($\gamma_m=1$), and $n=1$ one recovers $\Lambda$CDM with radiation. In what follows, we will be mainly interested in the role of the nonlinear term $\beta L_m^n$ as a possible driver of late-time acceleration. At the same time, $\gamma$ plays the role of a conventional vacuum contribution.

Focusing on cold matter ($p_m=0$) to isolate the role of $n$, we define
\bea
   \Omega_{m} \equiv  \frac{(2n-1)\rho^{n}_{m}}{3H^{2}}, \qquad 
   \Omega_{r} \equiv \frac{\rho_{r}}{3H^{2}}, \qquad 
   \Omega_{\gamma} \equiv -\frac{\gamma}{3H^{2}}, \label{EF_OmegasA}
\eea so that for $n\geq 1/2$ the dimensionless energy density of matter $\Omega_m$ is non-negative, while for $\gamma<0$ one has $\Omega_\gamma>0$ acting as an explicit vacuum term. We assume $\gamma<0$ (cf.\ \cite{jaybhaye2022cosmology, shukla2023dynamical}), and the Friedmann constraint becomes
\begin{equation}
    \Omega_m \;=\; 1 - \Omega_{r} - \Omega_{\gamma}. \label{EF1-new1}
\end{equation}
Introduce bounded variables
\be
   x \equiv \Omega_{r},  \qquad
   y \equiv \Omega_{\gamma}, \label{EN-vars}
\ee
with phase space $\Psi=\{(x,y):\, 0\le x\le 1,\;0\le y\le 1,\; x+y\le 1\}$. From the Friedmann–Raychaudhuri sector:
\be
    \Omega_m = 1 - x - y, \qquad
    2A \equiv 2\frac{\dot H}{H^{2}} = -3 - x + 3y + 3\frac{n-1}{2n-1}(1 - x - y). \label{EFs-new}
\ee
The deceleration parameter and the effective barotropic parameter take the form:
\bea
    q &=& -1-A \;=\; \frac{(5x-3y-1)n-4x+2}{4n-2}, \label{qxy}\\
    \omega_{\text{eff}} &=& -1-\tfrac{2}{3}A \;=\; \frac{(5x-3y-3)n-4x+3}{6n-3}. \label{weffxy}
\eea
The statefinder pair $(r,s)$ follows from $r=1+A'+2A^2+3A$ and $s=(r-1)/[3(q-\tfrac12)]$; the closed forms in terms of $(x,y,n)$ are simple, but long, so we will not present them explicitly.
We use
\begin{equation}
    c_a^2 \;=\; \omega_{\rm eff} \;-\; \frac{1}{3}\,\frac{\omega'_{\rm eff}}{1+\omega_{\rm eff}}
    \;=\; -1 - \frac{2}{3}A - \frac{A'}{3A}, \label{csa2-caseA}
\end{equation}
where $'\equiv d/d\ln a$ and $A'=A_x x'+A_y y'$ with $x'=-2x(2+A)$ and $y'=-2Ay$. At critical points $(x',y')=(0,0)$ one has $A'=0$, hence $c_a^2=\omega_{\rm eff}$ (with the de Sitter limit $A\to 0$ understood as a trajectory limit, $c_a^2\to -1$).

In terms of $\lambda=\ln a$, the dynamical system is given by:
\be
   x' \;=\; -2x\!\left(2+\frac{\dot{H}}{H^{2}}\right), \qquad 
   y' \;=\; -2y \left(\frac{\dot{H}}{H^{2}}\right). \label{SDxy}
\ee
The fixed points and diagnostics are:
\begin{itemize}
    \item \textbf{Matter} $P_m(0,0)$: $\Omega_m=1$, $q=\frac{-n+2}{4n-2}$, $\omega_{\rm eff}=\frac{-n+1}{2n-1}$, $c_a^2=\omega_{\rm eff}$ at the point, eigenvalues $\lambda_1 = -\frac{5n-4}{2n-1}$, $\lambda_2 = \frac{3n}{2n-1}$. Since the non-negativity of $\Omega_m$ requires that $2n-1>0$, the matter-dominated solution is an isolated source point (past attractor) if $n<4/5$, while it is a saddle point if $n>4/5$.
    
    \item \textbf{de Sitter} $P_\gamma(0,1)$: $\Omega_\gamma=1$, $q=-1$, $\omega_{\rm eff}=-1$, $c_a^2\to -1$, while the eigenvalues of the linearization matrix at the de Sitter solution read $\lambda_1=-4$, $\lambda_2=-\frac{3n}{2n-1}$. This is always a future attractor.
    
    \item \textbf{Radiation} $P_r(1,0)$: $\Omega_r=1$, $q=1$, $\omega_{\rm eff}=\tfrac13$, eigenvalues $\lambda_1=4$, $\lambda_2=\frac{5n-4}{2n-1}$. This is a saddle critical point if $n<4/5$ and a past attractor if $n>4/5$.
\end{itemize}

The qualitative behavior depends on $n$; a compact summary is given in Table~\ref{tab:alphabetaGen}. In brief: (i) for $n>4/5$ the sequence $\text{radiation}\to\text{matter}\to\text{de Sitter}$ is recovered; (ii) $n=\tfrac12$ is singular (decoupling limit); (iii) $0<n<\tfrac12$ is disfavored by the sign of $(2n-1)$ in the normalization and by the phase–space structure; (iv) $n<0$ admits late acceleration but departs from a standard matter era in geometric diagnostics. These trends are consistent with the observed values that cluster around $n\simeq 1$ reported in \cite{shukla2023dynamical, jaybhaye2022cosmology, garg2024cosmological, patil2023flrw}.

\begin{table}[h]
  \centering
  \resizebox{12cm}{!}{
  \begin{tabular}{|c|c|c|c|}
     \hline
     \textbf{Critical Point} & $\tfrac12<n<\tfrac45$ & $n>\tfrac45$ & $\Lambda$CDM ($n=1$) \\ 
     \hline \hline
     $P_m(0,0)$      & Past attractor  & Saddle         & Saddle  \\
     $P_r(1,0)$      & Saddle          & Past attractor & Past attractor \\ 
     $P_\gamma(0,1)$ & Future attractor & Future attractor & Future attractor \\ 
     \hline
  \end{tabular}
  }
  \caption{Stability of the fixed points of \eqref{SDxy} in the physically allowed domain $2n-1>0$ (hence $\Omega_m\ge 0$). The intervals $n\le 1/2$ and $n<0$ yield negative effective matter density and are therefore discarded; see the main text for a brief discussion.}
  \label{tab:alphabetaGen}
\end{table}

Values $n\le\tfrac12$ (including $n<0$) formally admit critical points but imply $\Omega_m<0$ through Eq. (\ref{EF_OmegasA}), so they are discarded from the physically viable sector. Our phase-space analysis below therefore focuses on $n>\tfrac12$.

\begin{figure}[h]
    \centering
    \includegraphics[width=5cm]{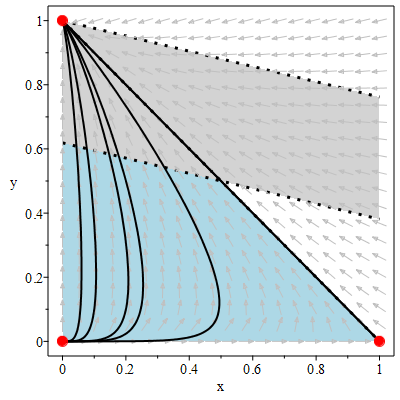}
    \includegraphics[width=5cm]{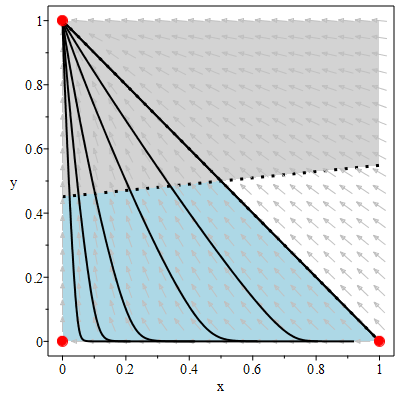}
    \includegraphics[width=5cm]{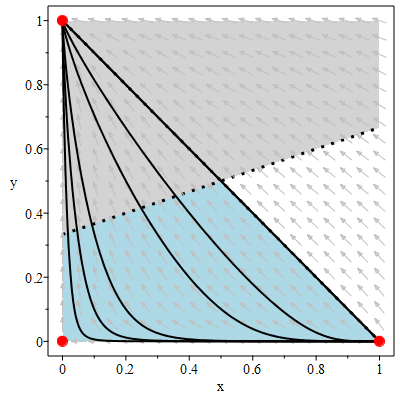}

    \caption{Phase-space portraits for the system (\ref{SDxy}) in the $(x,y)$ plane for Case~A. \textbf{Left:} representative value $\tfrac12<n<\tfrac45$ ($n=0.7$). \textbf{Center:} $n=0.85$ (near the transition $n=\tfrac45$). \textbf{Right:} $n=1$ ($\Lambda$CDM limit). Blue-shaded regions indicate physically viable matter densities ($0\leq\Omega_m\leq 1$). Gray-shaded regions correspond to $-1\leq q\leq 0$, representing accelerated expansion. Arrows denote the direction of evolution. Critical points are marked to highlight the dynamical behavior for different $n$.}
    \label{Fpoly1}
\end{figure}

Now, it is essential to consider the limit in which the explicit vacuum term is turned off, $\gamma=0$, so that $\Omega_\gamma\equiv 0$ and the phase space collapses to the one-dimensional subset $y=0$ with $0\leq x\leq 1$. In this case, the dynamics are entirely controlled by the nonlinear coupling $\beta L_m^n$ and by the radiation density component. The critical points $P_r(1,0)$ and $P_m(0,0)$ remain unchanged with respect to the whole system. The late-time fate of the Universe is determined by the sign and magnitude of $(2n-1)$: for $n>1/2$, a matter-dominated standard phase is recovered that does not approximate a de Sitter attractor in the absence of $\gamma$, whereas for $n<0$, accelerated expansion is possible but at the expense of a non-standard matter epoch, as reflected in the statefinder diagnostics. Therefore, in Case A, the term $\beta L_m^n$ alone is not sufficient to produce a viable late-time accelerated attractor compatible with a conventional matter epoch, which motivates the inclusion of an explicit vacuum contribution or the consideration of alternative model structures (as in Case B below).

Compared with previous $f(R,L_m)$ studies --- e.g.\ $f(R,L_m)=\Lambda+\tfrac{\alpha}{2}R+\beta L_m^n$ \cite{shukla2023dynamical} or $f(R,L_m)=\tfrac{R}{2}+L_m^n+\beta$ \cite{jaybhaye2022cosmology} --- our setup cleanly disentangles the explicit vacuum term ($\gamma$) from the nonlinear matter index ($n$), with the additional freedom encoded in $\beta$ controlling the overall strength of the nonlinear matter coupling. The dynamical analysis identifies the globally viable sector ($n\gtrsim 4/5$) that produces the standard sequence radiation $\to$ matter $\to$ de Sitter, and aligns with observational reconstructions that typically favor $n\approx 1$ \cite{garg2024cosmological, patil2023flrw}. Within this viable band, the proximity of $n$ to unity also parameterizes the timing of the onset of the de Sitter-like accelerated phase: for fixed $(\beta,\gamma)$, values of $n$ closer to $1$ drive an earlier departure from the matter-dominated regime, while smaller $n$ in the allowed interval delay the transition to acceleration, as reflected in the evolution of $q(z)$, $w_{\rm eff}(z)$ and in the statefinder trajectories.

The dynamical-systems analysis of Case~A shows that, once an explicit vacuum contribution is included ($\gamma\neq 0$), the late-time behavior is essentially fixed: the de Sitter point $P_\gamma(0,1)$ is always a global future attractor, independently of the value of $n$ within the physically admissible sector $2n-1>0$. Therefore, accelerated expansion is not generated as a genuine dynamical outcome of the nonlinear matter term, but rather enforced by the vacuum term $\gamma$, while the role of the exponent $n$ is mainly to control the duration and properties of the intermediate radiation/matter stages through the modified dilution law $\rho_m\propto a^{-3/(2n-1)}$. Requiring a standard radiation-to-matter transition and non-negative effective matter density restricts the phenomenologically viable range to $n\gtrsim 4/5$, for which the phase portrait closely resembles the $\Lambda$CDM sequence \emph{radiation} $\to$ \emph{matter} $\to$ \emph{de Sitter} (cf.\ Table \ref{tab:alphabetaGen} and Fig.~\ref{Fpoly1}).

In the complementary limit $\gamma=0$, the system collapses to $y=0$ and no accelerated late-time attractor compatible with a conventional matter epoch exists: the nonlinear term $\beta L_m^n$ alone is insufficient to reproduce a viable background history with radiation, matter domination, and asymptotic acceleration.
Taken together, these results indicate that Case A provides a useful consistency check and a controlled extension around $\Lambda$CDM, but offers limited new dynamical phenomenology.
This motivates exploring alternative $f(R,L_m)$ structures in which the nonlinear matter sector can serve as an effective dark-energy component with a nontrivial phase-space organization, as in Case B below.

\subsection{Case B: Nonlinear Matter Terms as Dynamical Dark Energy Mimickers} \label{SecIIIB}

In this section, we analyze the cosmological dynamics of the functional form
\begin{equation}
    f(R,L_{m}) = \frac{R}{2} + \beta \rho_{m} + \gamma \rho_{m}^{n},
\end{equation}
where $\beta$, $\gamma$, and $n$ are free parameters. 
From the unified template $f(R,L_m)=\alpha R/2 + c_1 L_m + c_n L_m^n + c_0$ with $\alpha =1$, $c_1=\beta$, $c_n=\gamma$, $L_m=\rho_m$, recovering standard GR with pressureless matter in the limit $n=1$ requires that the net coefficient of the linear matter term equals unity. Since $L_m=\rho_m$, this amounts to demanding $(\beta+\gamma)\rho_m \to \rho_m$, so that at the background level one must impose $\beta+\gamma=1$ (equivalently, $\beta=1-\gamma$). Once the Friedmann equations are written in terms of dimensionless density parameters, the effective matter sector depends only on the combination $(\beta+\gamma)$. Therefore, $\beta$ and $\gamma$ are not separately identifiable in the GR limit. For this reason, we keep both parameters explicit in the dynamical analysis and interpret constraint $\beta+\gamma=1$ as a consistency condition that can be enforced \textit{a posteriori} whenever recovery of standard GR at $n=1$ is required.

Guided by the previous case, here the linear term $\beta\rho_m$ plays the role of standard matter, while $\gamma\rho_m^n$ provides a genuinely nonlinear matter–curvature contribution that can mimic late–time acceleration.

Two particular limits of the model are worth emphasizing: $n=0$, where the nonlinear matter contribution reduces to a constant term $-\gamma$, which effectively plays the role of a cosmological constant, and $n=1$, where the nonlinear contribution merges with the standard matter density, effectively rescaling the matter sector as $(\beta+\gamma)\rho_m$. Thus, $n=0$ reproduces $\Lambda$CDM at the background level, whereas $n=1$ yields a trivial renormalization of the matter sector; departures occur for $n \neq 0,1$.

This formulation presents a novel perspective within $f(R,L_m)$ gravity, examining whether the combination of linear and nonlinear matter terms can account for late-time acceleration and thereby expand the landscape of viable cosmological models.

Using Eqs. (\ref{EF1}) and (\ref{EF2}), and assuming that matter and radiation behave as perfect fluids with $\omega_{m}=0$ and $\omega_{r}=1/3$, the Friedmann and continuity equations for the model take the form:
\bea \label{eq:friedmann_b}
    &&3H^{2} = \gamma (2n-1)\rho^{n}_{m} + \beta \rho_{m} + \rho_{r}, \label{2FE1} \\
    &&2 \dot H  + 3H^{2} =  \gamma  (n-1) \rho^{n}_{m} - \frac{1}{3}\rho_{r}, \label{2FE2} \\
    &&\left[ \gamma n (2n-1)\rho^{n-1}_{m} + \beta\right]\dot{\rho}_{m} + 3H\left(\gamma n \rho^{n}_{m} + \beta\rho_{m}\right) = 0, \label{2CM2} \\
    &&\dot{\rho}_{r} + 4H\rho_{r} = 0. \label{2CR2}
\eea  

From the first Friedmann equation (\ref{2FE1}) one has
\be
    1 = \frac{\gamma (2n-1)\rho^{n}_{m}}{3H^{2}}  + \frac{\beta \rho_{m}}{3H^{2}}  + \frac{\rho_{r}}{3H^{2}}. 
\ee
We define the normalized energy densities as
\be
   \Omega_{NL} \equiv \frac{\gamma (2n-1)\rho^{n}_{m}}{3H^{2}}, \qquad
   \Omega_{m} \equiv \frac{\beta \rho_{m}}{3H^{2}}, \qquad
   \Omega_{r} \equiv \frac{\rho_{r}}{3H^{2}}, \label{OsMB}
\ee
where $\Omega_{NL}$ denotes the nonlinear matter contribution, effectively acting as a dynamical dark-energy mimicker, $\Omega_m$ represents the standard linear matter density, and $\Omega_r$ corresponds to the radiation component. At the background level, only specific combinations of $(\beta,\gamma)$ enter the normalized densities, so that these coefficients can be reabsorbed into dimensionless amplitudes when confronting observations (see Sec. \ref{sec:obs} for the practical reparameterization used in the Bayesian analysis).

Introducing the bounded variables
\be
   x = \Omega_{r}, \qquad
   y = \Omega_{NL}, \label{nVars2}
\ee
the background equations read
\begin{equation}
    \Omega_m = 1 - x - y,  
    \qquad 
    \frac{\dot{H}}{H^{2}} = -\frac{1}{2} \left[ 3\Omega_m + 4x + \frac{3n}{(2n-1)}\, y \right]. \label{FEnV}
\end{equation}

Differentiating with respect to \(\lambda=\ln a\) (prime \( '\!=d/d\lambda\)),
\bea
    x' &=& -2x \left[ 2 + \frac{\dot{H}}{H^{2}} \right],  \label{DS-pol-linxA1} \\
    y' &=& -3ny \left[ \frac{\tfrac{ny}{2n-1}+\Omega_m}{ny+\Omega_m} \right] - 2y \frac{\dot{H}}{H^{2}}. \label{DS-pol-linyA1}
\eea

To ensure a physically viable phase space, we restrict ourselves to $\beta \geq 0$, $\gamma(2n-1)\geq 0$, and $\rho_m \geq 0$, so that all normalized densities are non-negative and satisfy $1 = x + \Omega_m + y$. Hence, the phase space is the simplex \(\Psi_{\rm phys}=\{(x,y)\,|\,0\le x\le1,\;0\le y\le1,\;x+y\le1\}\).

It is worth noting that this restriction formally requires $n \geq 1/2$. However, we consider $0<n<\tfrac12$ (with $\gamma<0$ ensuring $\Omega_{NL}\ge0$) and $n=0$ ($\Lambda$CDM limit).

Let \(A \equiv \dot H/H^{2}\). The cosmographic parameters are given by
\bea
    q &=& -1 - A = \frac{(2x-3y+2)n-x+3y-1}{4n-2}, \\
    \omega_{eff} &=& -1 - \tfrac{2}{3}A = \frac{(2x-3y)n-x+3y}{6n-3}, \\
    r &=& 1 + A' + 2A^{2} + 3A, \qquad A' \equiv \tfrac{dA}{d\lambda}, \qquad
    s =\frac{r-1}{3\left(q-\tfrac{1}{2}\right)}.
\eea

For the background mixture, we use the adiabatic diagnostic
$c_a^2=\omega_{\rm eff}-\frac{1}{3}\frac{\omega'_{\rm eff}}{1+\omega_{\rm eff}}$, with $\omega'_{\rm eff}=\omega_{{\rm eff},x}x'+\omega_{{\rm eff},y}y'$.
At fixed points $x'=y'=0$ one has $c_a^2=\omega_{\rm eff}$; in de Sitter limits $A\!\to\!0$, $c_a^2\to-1$ along regular approaches.

The critical points $P_{i}:(x_{i},y_{i})$ within $\Psi_{\rm phys}$ are:

\begin{itemize}
    \item \textbf{Radiation (linear)}, $P_r(1,0)$: $\Omega_m=0$, $\Omega_r=1$, $\Omega_{NL}=0$; $q=1$, $\omega_{\rm eff}=\tfrac{1}{3}$, $(r,s)=(3,\tfrac{4}{3})$. The linearization matrix evaluated at the radiation-dominated critical point has the eigenvalues $\lambda_1=1$, $\lambda_2=4-3n$, so $P_r$ is a past attractor for $n<4/3$ and becomes non-hyperbolic at $n=4/3$. On the invariant edge $\Omega_m=0$, the one-dimensional tangent eigenvalue (governing perturbations confined to this edge) is $\sigma_r=\frac{5n-4}{2n-1}$.

    \item \textbf{Matter (linear)}, $P_m(0,0)$:
    $\Omega_m=1$, $\Omega_r=0$, $\Omega_{NL}=0$. 
    $q=\tfrac{1}{2}$, $\omega_{\rm eff}=0$, $(r,s)=(1,\text{undefined})$. Eigenvalues: $\lambda_1=-1,\;\lambda_2=3(1-n)$. Saddle for $n<1$; future attractor for $n\ge 1$.

    \item \textbf{Nonlinear contribution}, $P_{NL}(0,1)$:
    $\Omega_m=0$, $\Omega_r=0$, $\Omega_{NL}=1$. 
    $q=\frac{2-n}{4n-2}$, $\omega_{\rm eff}=\frac{1-n}{2n-1}$,
    $(r,s)=\big(\frac{-n^2+n+2}{2(2n-1)^2},\,\frac{n}{2n-1}\big)$. Eigenvalues: $\lambda_1=-\frac{5n-4}{2n-1},\;
    \lambda_2=\frac{3(n-1)}{2n-1}$. Future attractor only for $\tfrac{4}{5}<n<1$; otherwise, saddle.
    
    \item \textbf{de Sitter (scaling)}, $P_{dS}\!\left(0,\frac{2n-1}{\,n-1\,}\right)$:
    $\Omega_r=0$, $\Omega_{NL}=\frac{2n-1}{n-1}$, $\Omega_m=\frac{n}{1-n}$. $q=-1$, $\omega_{\rm eff}=-1$, $(r,s)=(1,0)$. Eigenvalues $\lambda_1=-4$, $\lambda_2=-\tfrac{3}{2}$. This solution exists; that is, it is a physical solution for $0\leq n<\tfrac12$, where it exists, it is the future attractor.
\end{itemize}

\begin{table}[h]
\centering
\resizebox{16cm}{!}{
\begin{tabular}{|c|c|c|c|c|c|c|c|}
\hline
\textbf{Critical Point} & \textbf{$n<0$} & \textbf{$n=0$} & \textbf{$0 < n < 1/2$} & \textbf{$1/2 < n \leq 4/5$} & \textbf{$4/5 < n < 1$} & \textbf{$1 < n \leq 4/3$} & \textbf{$n > 4/3$} \\ 
\hline
$P_m(0,0)$ & Saddle & Saddle & Saddle & Saddle & Saddle & Future attractor & Future attractor \\ 
\hline
$P_{NL}(0,1)$ & Saddle & Saddle & Saddle & Saddle & Future attractor & Saddle & Saddle \\ 
\hline
$P_r(1,0)$ & Past attractor & Past attractor & Past attractor & Past attractor & Past attractor & Past attractor & Past attractor \\ 
\hline
$P_{dS}\left(0,\tfrac{2n-1}{n-1}\right)$ & Non-physical & Non-physical & Future attractor & Non-physical & Non-physical & Non-physical & Non-physical \\ 
\hline
\end{tabular}
}
\caption{Stability of the critical points of the system (\ref{DS-pol-linxA1})--(\ref{DS-pol-linyA1}) as a function of $n$. The de Sitter point $P_{dS}$ lies in the physical domain only for $0<n<1/2$.}
\label{CaseBstability}
\end{table}

Thus, \(0<n<\tfrac12\) is the robust branch: \(P_r\!\to\!P_m\!\to\!P_{dS}\) with background stability (at the fixed points \(c_a^2=\omega_{\rm eff}\); near de Sitter \(c_a^2\to-1\)). Physical stability (no–ghost/no–gradient, \(c_s^2\), \(\mu,\Sigma,\eta\)) requires the perturbative analysis to be deferred to a companion work.

\begin{figure}[h]
    \centering
    \includegraphics[width=5cm]{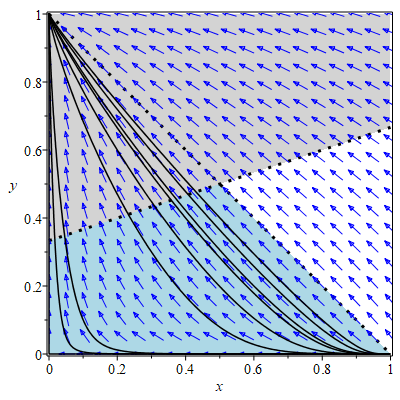}
    \includegraphics[width=5cm]{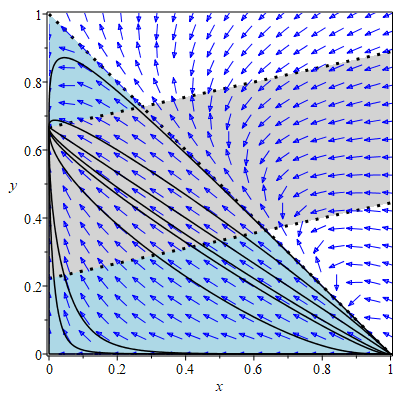}
    \includegraphics[width=5cm]{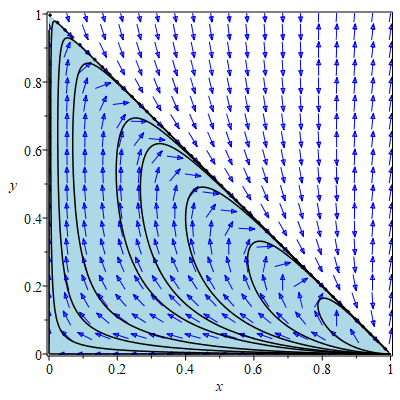}
    \includegraphics[width=5cm]{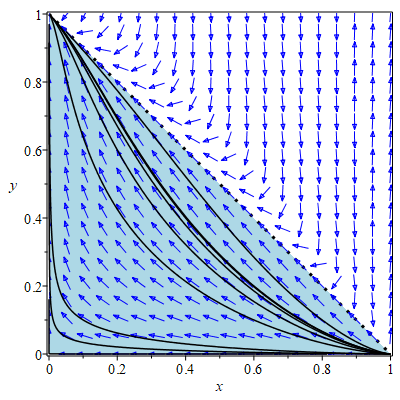}
    \includegraphics[width=5cm]{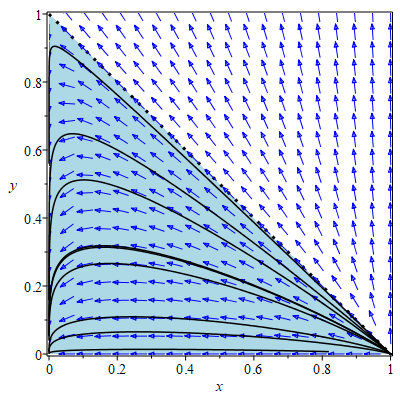}
    \includegraphics[width=5cm]{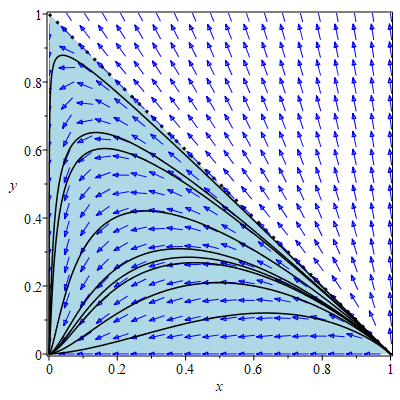}
   \caption{Phase–space portraits of the system (\ref{DS-pol-linxA1})–(\ref{DS-pol-linyA1}) for representative $n$. Top-left: $n=0$ ($\Lambda$CDM limit). Top-center: $n=0.15$ (viable accelerating regime $0<n<1/2$). Top-right: $n=0.75$ (non-accelerating regime $1/2<n<4/5$). Bottom-left: $n=0.85$ (nonlinear attractor regime $4/5<n<1$). Bottom-center: $n=1.1$ (future matter attractor, $1<n\le 4/3$). Bottom-right: $n=1.5$ (asymptotic matter–dominated attractor, $n>4/3$).}
   \label{FP+L1}
\end{figure}

The qualitative dynamics of the system are strongly determined by the value of the parameter $n$:

\begin{itemize}
    \item $n<0$: In this inverse–power regime, the nonlinear term grows as $\rho_m$ dilutes. The radiation point $P_r(1,0)$ is a past attractor (source), while the linear–matter point $P_m(0,0)$ is a saddle, and the nonlinear corner $P_{NL}(0,1)$ is also a saddle. The de Sitter equilibrium $P_{dS} \left(0,\frac{2n-1}{n-1}\right)$ lies outside the bounded physical simplex ($x\ge0,\,y\ge0,\,x+y\le1$; effectively $y>1$), so there is no late–time accelerating attractor in the physical domain. See, \textit{e.g.}, inverse–density/phantom–like constructions for conceptual parallels \citep{caldwell2002phantom,bamba2009crossing}. 

    \item $0<n<\tfrac{1}{2}$: This interval constitutes the physically most relevant regime of the model, as it admits a complete cosmic sequence within the bounded phase space. The radiation point $P_r$ acts as a past attractor, while the matter point $P_m$ is a saddle, allowing for a transient matter-dominated epoch consistent with structure formation. The nonlinear point $P_{NL}$ is also a saddle, which shows that the nonlinear density term alone cannot sustain a stable cosmological phase. The de Sitter point $P_{dS}$ lies within the physical domain and emerges as a global future attractor, ensuring asymptotic convergence towards accelerated expansion. Unlike $\Lambda$CDM—where the attractor is vacuum—here acceleration is driven by a scaling de Sitter state with fixed ($\Omega_m,\Omega_{NL}$). This reproduces the sequence radiation $\to$ matter $\to$ acceleration without an explicit cosmological constant (cf. Fig.~\ref{FP+L1}).

    \item $\tfrac{1}{2}<n<\tfrac{4}{5}$: The global phase–space structure changes qualitatively with respect to $0<n<\tfrac{1}{2}$. The point $P_{NL}$ remains a saddle (with reversed eigenvalue signs along $x$/$y$), and $P_{dS}$ ceases to exist within the physical simplex, eliminating a stable late–time accelerating attractor. Trajectories are transient (radiation/matter–like) and do not converge to acceleration (see Fig.~\ref{FP+L1}).

    \item $\tfrac{4}{5}<n<1$: The nonlinear point $P_{NL}$ becomes a hyperbolic sink (global future attractor). $P_r$ remains a past attractor and $P_m$ a saddle. However, the final state is non–accelerating since $q>0$ and $\omega_{\rm eff}>-1/3$ at $P_{NL}$. Thus, the sequence radiation $\to$ matter $\to$ nonlinear domination fails to yield late–time acceleration (Fig.~\ref{FP+L1}).

    \item $1<n\leq \tfrac{4}{3}$: $P_r$ remains a past attractor, while $P_m$ becomes a future attractor, leading asymptotically to a purely matter–dominated state. $P_{NL}$ is a saddle and $P_{dS}$ lies outside the physical domain. No viable late–time acceleration (Fig.~\ref{FP+L1}).

    \item $n>\tfrac{4}{3}$: $P_r$ is a past attractor and $P_m$ the global future attractor. $P_{NL}$ remains a saddle and $P_{dS}$ is non–physical. The sequence is radiation $\to$ matter with no mechanism for late–time acceleration, and therefore observationally disfavored (Fig.~\ref{FP+L1}).
\end{itemize}

Among the regimes summarized in Table \ref{CaseBstability}, only $0<n<\tfrac12$ simultaneously admits a standard radiation– and matter–dominated era followed by a physically allowed de Sitter–like accelerated attractor within the bounded simplex $\Psi_{\rm phys}$. In the remaining intervals, either the would–be de Sitter point lies outside the physical domain or the late–time attractor is non-accelerating. Therefore, in what follows, we regard $0<n<\tfrac12$ as the phenomenologically viable window of Case~B, and we restrict the observational analysis to this branch.

We provide two quick, background–level estimators for the exponent $n$ under the assumptions of spatial flatness, negligible present radiation ($\Omega_{r0}\!\ll\!1$), and current background constraints \citep{aghanim2020planck}. These estimates are indicative and do not replace a full likelihood analysis.

\begin{enumerate}
\item Assuming the Universe sits at the scaling de Sitter attractor ($0<n<\tfrac12$).
At $P_{\rm dS}$ one has $\Omega_{m}=\dfrac{n}{1-n}$. Hence
$n=\frac{\Omega_{m0}}{1+\Omega_{m0}}$. With $\Omega_{m0}=0.315\pm0.007$  \citep{aghanim2020planck} one obtains
\be
    n=0.240\pm 0.004.
\ee

\item Without assuming the attractor (using $q_0,\Omega_{m0}$).
From 
$q=-1+\frac{1}{2}\big[\,3\Omega_m+4\Omega_r+\frac{3n}{2n-1}\Omega_{NL}\big]$
and $\Omega_m+\Omega_r+\Omega_{NL}=1$, eliminating $\Omega_{NL}$ yields
\be
    n=\frac{3\Omega_{m0}+4\Omega_{r0}-2q_0-2}{3\Omega_{m0}+5\Omega_{r0}-4q_0-1}.
\ee
Taking $\Omega_{r0}\!\ll\!1$, $\Omega_{m0}=0.315\pm0.007$ and
$q_0\simeq -0.528\pm0.011$ (as inferred under $\Lambda$CDM from Planck 2018 \citep{aghanim2020planck}),
\be
    n \approx (4.9\times 10^{-4}) \pm 0.015,
\ee
which is statistically consistent with $n=0$ ($\Lambda$CDM kinematics at background level).
\end{enumerate}

The second estimator is slightly model-dependent, since $q_0$ is obtained from a reconstruction of $\Lambda$CDM; therefore, it is conservative to present it as an indicative consistency check. A full joint fit (SNIa+CC+BAO+CMB) is required to determine whether we are currently in the scale attractor (suggested by $n\simeq0.24$) or whether we are still following the $\Lambda$CDM path ($n\simeq 0$).

The autonomous system in the original variables $(x,y)$ may send critical configurations to infinity when $y$ diverges, notably near the singular value $n=\tfrac12$ where factors $(2n-1)$ vanish. To verify global robustness, we also analyze a compactified phase space via $v\equiv y/(1+y)\in[0,1)$. The compact analysis recovers the same qualitative structure: the interval $0<n<\tfrac12$ uniquely admits a physical de Sitter \emph{scaling} attractor (radiation $\to$ matter $\to$ acceleration), whereas for $n\ge\tfrac12$ the de Sitter point exits the physical domain and only saddle or matter–dominated late-time states remain. Thus, compactification confirms that our viability window is restricted to $0<n<\tfrac12$.

The phase–space analysis above singles out a genuinely distinctive and phenomenologically appealing branch of the theory: for $0<n<\tfrac12$, the Universe evolves through the standard sequence radiation $\rightarrow$ matter $\rightarrow$ accelerated expansion without invoking an explicit vacuum term. In this window, the late–time de Sitter state is not a pure--$\Lambda$ attractor but a scaling fixed point, in which acceleration is generated dynamically by the nonlinear matter sector ($\Omega_{NL}\neq 0$) and persists as a stable future sink within the bounded physical simplex. This sharply contrasts with Case A, where a viable accelerated attractor essentially relies on an explicit vacuum contribution and the dynamics largely reduce to a $\Lambda$CDM-like structure. Consequently, Case B provides a concrete mechanism for late–time acceleration emerging from a curvature--matter coupling encoded in $\gamma\rho_m^{\,n}$, and the key question becomes quantitative: whether the viable range $0<n<\tfrac12$ (and its predicted scaling composition) can reproduce the observed expansion history and distance indicators. This motivates the Bayesian observational fit performed in Sec. \ref{sec:obs}, where we confront the accelerating branch of Case B with SNIa+CC+BAO data and assess parameter identifiability and viability beyond the background fixed-point diagnostics.

From a physical viewpoint, the nonlinear contribution $\rho_m^n$ in the viable window $0<n<\tfrac12$ is most consistently interpreted as an effective term induced by the non-minimal matter--geometry coupling, rather than as a new fundamental matter species in the universe. Standard perfect-fluid components that obey the usual conservation law $\rho\propto a^{-3(1+w)}$ do not, by themselves, generate fractional powers of the same density in the background energy budget. In the present setup, the term $\gamma\rho_m^n$ should therefore be viewed as a density-dependent interaction/response of the coupled sector: for $n<1$ it dilutes more slowly than $\rho_m$ and can naturally dominate at late times, thereby mimicking dark-energy behavior without introducing an explicit vacuum contribution. While possible microphysical completions (effective-medium descriptions, coarse-graining/backreaction corrections, or interaction-induced composites) are model-dependent and not assumed here, throughout this work $\rho_m^n$ is treated as a phenomenological proxy capturing the leading density dependence of the coupled sector at the background level. This motivates, as a next step, a minimal consistency check of stability before confronting the model with background observables.

Let us start with Case B. As said, it is useful to record minimal consistency requirements regarding stability. In metric theories with an effective Einstein--Hilbert sector, tensor perturbations propagate with a kinetic prefactor proportional to the effective Planck mass, $M_*^2\propto f_R$, so the absence of tensor ghost instabilities requires $f_R>0$ along the cosmological evolution. In the present realizations, $f(R,L_m)=R/2+\cdots$ implies
\begin{equation}
    f_R=\frac{1}{2}=\mathrm{const.}>0,
\end{equation}
hence the tensor sector is ghost-free at background level, and the effective gravitational coupling remains positive (no sign flip of $G_{\rm eff}$). Moreover, because the model contains no derivative couplings that modify the principal part of the tensor equation, gravitational waves propagate luminally as in GR, \textit{i.e.}, $c_T^2=1>0$, so there are no tensor Laplacian (gradient) instabilities.

In addition to the tensor sector, the matter--curvature interaction is controlled by $f_{L_m}$, which for Case B reads as:
\begin{equation}
    f_{L_m}=\beta+\gamma n\,\rho_m^{\,n-1}.
\end{equation}
Requiring $f_{L_m}>0$ over the relevant density range avoids pathological sign reversals in the effective matter--geometry coupling and is compatible with the physically restricted phase space adopted above (together with $\rho_m\ge0$ and the sign choice ensuring $\Omega_{NL}\ge0$). Finally, within the accelerating branch $0<n<\tfrac12$ singled out by the phase--space analysis, trajectories converge to the scaling de Sitter attractor; along regular approaches one has $A\to0$ and $c_a^2\to-1$, which supports background--level adiabatic consistency near the late--time attractor.

We emphasize that these criteria are \emph{necessary but not sufficient} for full viability. A definitive assessment requires the complete linear perturbation analysis, including scalar no--ghost and no--gradient conditions (positivity of the kinetic term and sound speed), as well as the computation of the effective modified--gravity functions $\mu(k,a)$ and $\Sigma(k,a)$ and the gravitational slip $\eta(k,a)$, to be confronted with growth and lensing data. Since the present work focuses on background dynamics, global phase--space structure, and Bayesian constraints from SNIa+CC+BAO(+CMB), we defer this exhaustive perturbative analysis to a future paper, where stability conditions will be implemented across the posterior--supported parameter space and the model will be tested against structure--formation observables. The possible impact of the models on the $H_0$-tension issue will also be explored.


\section{Observational analysis}\label{sec:obs}


We confront the background expansion of the $f(R,L_m)$ models with late-- and intermediate--redshift probes and benchmark the results against the $\Lambda$CDM model.
In the dynamical-systems analysis of Secs. \ref{SecIIIA}--\ref{SecIIIB}, the nonrelativistic sector was represented by a single effective fluid with density $\rho_m$. 
Throughout we use units $8\pi G=c=1$, so that $[R]=[\rho_m]\sim H^2$. In this normalization, the couplings $(\beta,\gamma)$ entering $f(R,L_m)$ should be regarded as effective parameters defined after fixing a reference density scale.
Importantly, $\gamma$ does not carry the same mass dimension in the two realizations: in Case A, $f(R,L_m)=\tfrac{R}{2}+\beta \rho_{dm}^{\,n}+\gamma$, the constant term $\gamma$ is vacuum--energy--like and therefore has dimension $[\gamma]\sim H^2$,
whereas in Case~B, $f(R,L_m)=\tfrac{R}{2}+\beta\rho_{dm}+\gamma\rho_{dm}^{\,n}$, the coefficient of the linear matter term is dimensionless and the nonlinear amplitude is most transparently encoded in the present-day density fraction $\Omega_{NL,0}$.
Consequently, for background dynamics and parameter inference, it is convenient to trade $(\beta,\gamma)$ for dimensionless density amplitudes that enter the normalized Friedmann equation and the likelihood. Concretely, we sample $(\Omega_{dm,0},\Omega_{b,0},h,n)$ (with a BBN prior on $\Omega_b h^2$), while the remaining amplitude is derived from spatial flatness,
namely $\Omega_{\gamma,0}=1-\Omega_{dm,0}-\Omega_{b,0}-\Omega_{r,0}$ in Case~A and $\Omega_{NL,0}=1-\Omega_{dm,0}-\Omega_{b,0}-\Omega_{r,0}$ in Case~B.
When needed, the original couplings can be reconstructed from these amplitudes (e.g.\ $\gamma=-3H_0^2\Omega_{\gamma,0}$ in Case~A, and $\gamma$ related to $\Omega_{NL,0}$ through Eq. \eqref{eq:gamma_caseB_omega} after fixing the normalization of $\beta$).

For observational analysis, it is convenient to split the total matter sector as $\rho_m=\rho_{dm}+\rho_b$, where $\rho_{dm}$ denotes the (pressureless) dark-matter component entering the matter Lagrangian $L_m$ and thus subject to the modified continuity equation implied by the curvature--matter coupling, while $\rho_b$ denotes the baryon contribution, assumed to be minimally coupled and to satisfy the standard conservation law, $\rho_b\propto a^{-3}$.
At the background level, this split does not modify the Friedmann-form description of the expansion history.
Still, it changes the bookkeeping of present-day density parameters used in the likelihood, i.e., $\Omega_{m0}=\Omega_{dm,0}+\Omega_{b,0}$.
Strictly speaking, treating $\Omega_b$ as an independent conserved component enlarges the autonomous system by one dimension; we keep the dynamical analysis in terms of the coupled effective sector and use the split only for the observational fit through the relation $\Omega_{m0}=\Omega_{dm,0}+\Omega_{b,0}$.

\textbf{Case A: Nonlinear Matter Coupling with Vacuum Term.}

\begin{equation}
    f(R,L_{dm})= \frac{R}{2}+ \beta \rho_{dm}^{\,n}+\gamma.
\end{equation} At the background level, we parametrize the model by $(\Omega_{dm,0},\Omega_{b,0},h,n)$. The vacuum contribution is fixed by spatial flatness,
$\Omega_{\gamma,0}=1-\Omega_{dm,0}-\Omega_{b,0}-\Omega_{r,0}$, so that $\gamma$ is treated as a derived parameter. To keep dimensionless amplitudes for arbitrary $n$, we define

\begin{equation}
    \Omega_{dm}(z) \equiv  \frac{\beta(2n-1)}{3H^{2}} \left( \frac{\rho_{dm}}{\rho_{dm,0}}\right)^n\rho_{dm,0},
    \qquad
    \Omega_{\gamma}(z) \equiv -\frac{\gamma}{3H^{2}},
    \label{EF_Omegas_obs}
\end{equation} and using the modified dark-matter continuity equation (Sec. \ref{SecIIIA}), one obtains an analytic solution for $\rho_{dm}(z)$, which leads to the modified Hubble function

\begin{equation}
    E^2(z)\equiv \frac{H^2(z)}{H_0^2}
    = \Omega_{dm,0} (1+z)^{\frac{3n}{2n-1}}
      + \Omega_{b,0}(1+z)^3
      + \Omega_{r,0}(1+z)^4
      + \Omega_{\gamma,0},
\end{equation} where flatness implies $\Omega_{\gamma}=1-\Omega_{dm}-\Omega_{b}-\Omega_{r}$ and therefore

\begin{equation}
    \gamma=-3H_0^2\,\Omega_{\gamma,0}=-3H_0^2\left(1-\Omega_{dm,0}-\Omega_{b,0}-\Omega_{r,0}\right).
\end{equation}


\textbf{Case B: Nonlinear Matter Terms as Dynamical Dark Energy Mimickers.}
\\

We consider the curvature--matter coupling

\begin{equation}
    f(R,L_{dm}) = \frac{R}{2}+ \beta \rho_{dm} + \gamma \rho_{dm}^{n},
\end{equation} where $\rho_{dm}$ denotes the pressureless dark-matter density entering the matter Lagrangian $L_{dm}$, while baryons are assumed to be minimally coupled and follow $\rho_b\propto a^{-3}$.

To work with dimensionless density parameters for arbitrary $n$, we define

\begin{equation}
  \Omega_{dm}(z) \equiv \frac{\beta \rho_{dm}}{3 H^2}, 
  \qquad
  \Omega_{NL}(z) \equiv \frac{\gamma (2 n - 1)}{3 H^2}\,
  \left( \frac{\rho_{dm}}{\rho_{dm,0}}\right)^n \rho_{dm,0},
  \label{eq:def1_obs}
\end{equation} so that the (spatially flat) Friedmann constraint reads

\begin{equation}
    1=\Omega_{dm}(z)+\Omega_{b}(z)+\Omega_{r}(z)+\Omega_{NL}(z).
\end{equation} With these conventions, $\Omega_{dm}$ behaves as the standard linear dust contribution, whereas $\Omega_{NL}$ encodes the nonlinear sector induced by the $\gamma\rho_{dm}^n$ coupling.

Introducing the normalized dark-matter density
$u(z)\equiv \rho_{dm}(z)/\rho_{dm,0}$ (with $u(0)=1$), the modified dark-matter continuity equation can be written as

\begin{equation}
  \frac{du}{dz} = \frac{3}{1+z}\frac{u + \dfrac{n \Omega_{NL,0}}{\Omega_{dm,0}(2n-1)}u^n}{1+\dfrac{n\,\Omega_{NL,0}}{\Omega_{dm,0}}\,u^{\,n-1}},  \qquad  u(0)=1,
  \label{eq:u_evolution_caseB}
\end{equation} where $\Omega_{dm,0}\equiv \Omega_{dm}(0)$ and $\Omega_{NL,0}\equiv \Omega_{NL}(0)$.

The Hubble rate is then obtained as

\begin{equation}
    E^2(z)\equiv \frac{H^2(z)}{H_0^2} = \Omega_{dm,0} u(z) + \Omega_{b,0}(1+z)^3 +\Omega_{r,0}(1+z)^4 + \Omega_{NL,0} u^n(z), \label{eq:E2_caseB}
\end{equation} with $\Omega_{NL,0}$ fixed by flatness as
$\Omega_{NL,0}=1-\Omega_{dm,0}-\Omega_{b,0}-\Omega_{r,0}.$

From the definition \eqref{eq:def1_obs} evaluated at $z=0$ (i.e. $u=1$),

\begin{equation}
\Omega_{NL,0} = \frac{\gamma(2n-1)}{3H_0^2} \rho_{dm,0}.
\end{equation}
Using $\Omega_{dm,0}=\beta\rho_{dm,0}/(3H_0^2)$, this can be equivalently written as
\begin{equation}
    \gamma = \frac{\beta \Omega_{NL,0}}{\Omega_{dm,0}(2n-1)}.
    \label{eq:gamma_caseB_omega}
\end{equation}
Therefore, after fixing $\beta=1$, $\gamma$ becomes a derived parameter:
\begin{equation}
    \gamma = \frac{\Omega_{NL,0}}{\Omega_{dm,0}(2n-1)} \qquad (\beta=1).
\end{equation}

For both cases, the departure from $\Lambda$CDM at the background level is governed by the nonlinear exponent $n$ in the curvature--matter sector. The remaining quantities entering the likelihood are standard late-time parameters, namely the total matter density $\Omega_m=\Omega_{dm,0}+\Omega_{b,0}$, the physical baryon density $\Omega_b h^2$, and the reduced Hubble constant $h\equiv H_0/(100\,{\rm km\,s^{-1}\,Mpc^{-1}})$. In our parameter estimation, we treat $n$ as the only additional (non-$\Lambda$CDM) degree of freedom. At the same time, the effective vacuum/nonlinear amplitudes are fixed by spatial flatness, i.e. $\Omega_{\gamma,0}=1-\Omega_{dm,0}-\Omega_{b,0}-\Omega_{r,0}$ in Case A and $\Omega_{NL,0}=1-\Omega_{dm,0}-\Omega_{b,0}-\Omega_{r,0}$ in Case B.

\subsection{Datasets}

Throughout the observational analysis, we used the collection of 31 cosmic chronometers (\textbf{CC}) \citep{Jimenez:2003iv, Simon:2004tf, Stern:2009ep, Moresco:2012by, Zhang:2012mp, Moresco:2015cya, Moresco:2016mzx}; the dataset (and the corresponding covariance information used in our analysis) is available in the public repository \citep{hz}. In combination with the CC, we include an up-to-date `Union' compilation of 2,087 SNe Ia from 24 datasets (\textbf{Union3}). The information of these 2,087 SNe Ia is compressed in 22 non-zero nodes, available in \citep{UNION3}, where the redshift nodes are in steps of 0.05 up to $z=0.8$, and then in steps of 0.1 (or every 10 SNe, whichever is larger in redshift).
In addition to CC and Union3, we also incorporate \textbf{DESI BAO} measurements from the first DESI cosmology analyses. For galaxies and quasars, we use the BAO determinations obtained from the Bright Galaxy Sample ({BGS}), spanning $0.1<z<0.4$; the Luminous Red Galaxy sample ({LRG}) in the intervals $0.4<z<0.6$ and $0.6<z<0.8$; the Emission Line Galaxy sample ({ELG}) covering $1.1<z<1.6$; the combined {LRG+ELG} sample in $0.8<z<1.1$; and the Quasar sample ({QSO}) over $0.8<z<2.14$. Beyond the galaxy BAO, we include BAO constraints from the Lyman-$\alpha$ forest ({Ly-$\alpha$}), spanning $1.77<z<4.16$, as reported in the dedicated DESI analysis \citep{DESI:2024mwx}. 
In addition to these datasets, we impose a Gaussian prior on $\Omega_b h^2$, motivated by big bang nucleosynthesis constraints \citep{Cooke:2017cwo}.

\subsection{Parameter Estimation}

We explore the parameter space within the Bayesian-inference framework. The posterior probability distribution of the model parameters $\theta$,
$\mathcal{P}(\theta \mid D,M)$, is obtained by combining the likelihood function $\mathcal{L}(D\mid\theta,M)$ with the prior distribution
$\pi(\theta\mid M)$, namely
\begin{equation}
    \mathcal{P}(\theta\mid D,M) =
    \frac{\mathcal{L}(D\mid\theta,M)\,\pi(\theta\mid M)}{\mathcal{Z}(D\mid M)},
\end{equation}
where $D$ denotes the observational data and $\mathcal{Z}(D\mid M)$ is the Bayesian evidence of the model~$M$. 
Assuming Gaussian-distributed observational errors, the likelihood can be written as
$\mathcal{L}\propto \exp\!\left(-\chi^2/2\right)$ (up to an additive constant in $-2\ln\mathcal{L}$), with
\begin{equation}
    \chi^2_{\rm data} =
    (d_{m}-d_{\rm obs})^{T}\,C_{\rm data}^{-1}\,(d_{m}-d_{\rm obs}),
\end{equation}
where $d_{m}$ and $d_{\rm obs}$ are the model predictions and observed data vectors, respectively, and $C_{\rm data}$ is the covariance matrix associated with each dataset. Since the set of observations considered throughout the analysis is independent of each other, the joint $\chi^2$ can be computed as
\begin{equation}
    \chi^2_{\rm total} = \chi^2_{\rm CC} + \chi^2_{\rm SN} + \chi^2_{\rm BAO}.
\end{equation}

 Regarding the priors used for the model parameters, we assume uniform (flat) priors with the following ranges: $\Omega_{\rm m} = [0.1, 0.5]$ for the matter density parameter, $\Omega_{\rm b} h^2 = [0.02, 0.025]$ for the physical baryon density, $h = [0.4, 0.9]$ for the dimensionless reduced Hubble constant (where $h = H_0 /100 \, \text{km} \, \text{s}^{-1} \, \text{Mpc}^{-1} $); according to the analysis of Sec. \ref{SecIIIA} we have $n = [4/5, 1.5]$ for case A and, similarly from Sec. \ref{SecIIIB} $n = [-0.3, 1/2]$ for case B. 
\\

By examining the best-fit parameters and their uncertainties, one can identify potential discrepancies that may indicate systematic errors, dataset tensions, or the need to review the underlying cosmological framework. To assess the relative performance of the theoretical scenario most favored by the data, a model comparison criterion is required \cite{liddle2007information}.
The Bayesian evidence can be used to compare two distinct models by naturally penalizing the complexity and additional parameters, by using the following relationship:
\begin{equation}
    B_{12} \equiv \frac{\mathcal{Z}(D|M_1)}{\mathcal{Z}(D|M_2)},
\end{equation}
where $B_{12}$ is known as the \textit{Bayes Factor}, with $M_1$ and $M_2$ being the two models to be compared. By taking the natural logarithm of this quotient, $\Delta\ln{\mathcal{Z}}$, together with the Jeffreys' scale, shown in Table \ref{tab:jeffreys}, \citep{kass1995bayes, efron2001scales, trotta2008bayes}, we have a useful empirical interpretation to perform model selection. Therefore, the Bayesian evidence can be used to assess how well the model $M_1$ compares with the model $M_2$.

However, computing the Bayesian evidence can be challenging in some settings, as it may require evaluating too many dimensions.  A simpler, and complementary, model selection technique is to compare goodness-of-fit and complexity, for instance, the Akaike \cite{akaike2003new} and Bayesian \cite{schwarz1978estimating} Information Criteria (AIC and BIC, respectively), which evaluate the maximum likelihood estimate and penalize model complexity. The AIC penalizes the addition of free parameters $k$, while the BIC also accounts for the number of data points $N$, as follows:
\begin{equation}
    {\rm AIC}=-2\ln \mathcal{L}_{\text{max}} + 2k,
    \label{AIC_criteria}
\end{equation}
\begin{equation}
    {\rm BIC}=-2\ln \mathcal{L}_{\text{max}}+ k \ln N.
    \label{BIC_criteria}
\end{equation}
Lower values of AIC and BIC indicate a more favored model under the corresponding criterion.

\begin{table}[t]
\centering
\begin{tabular}{ccc}
\hline
$\Delta\ln\mathcal{Z}\equiv \ln(\mathcal{Z}_1/\mathcal{Z}_2)$
& Odds $\mathcal{Z}_1:\mathcal{Z}_2$
& Evidence in favor of $M_1$ \\
\hline
$<1$ & $<3:1$ & Inconclusive \\
$[1,\,2.5)$ & $3:1$ to $12:1$ & Weak (or “substantial”) \\
$[2.5,\,5)$ & $12:1$ to $150:1$ & Moderate \\
$\ge 5$ & $\ge 150:1$ & Strong \\
\hline
\end{tabular}
\caption{Jeffreys-style interpretation of the Bayes factor in terms of the log-evidence difference $\Delta\ln\mathcal{Z}$.}
\label{tab:jeffreys}
\end{table}

To determine the optimal parameter values of the model, along with its statistics, we incorporated into the \texttt{SimpleMC} code~\cite{BOSS:2014hhw}, an adapted version of \texttt{dynesty}, a dynamic nested sampling library which efficiently computes the Bayesian evidence and bypasses the parameter estimation (see Ref. \citep{simplemc_github, Speagle:2019ivv}). SimpleMC is designed for parameter estimation and model comparison using Bayesian inference, optimization, and machine learning algorithms; more information on the application of this code can be found in the Bayesian inference review presented in~\citep{Padilla:2019mgi}.

\subsection{Results}

\begin{figure}[h]
    \centering
     \makebox[12cm][c]{
    \includegraphics[width=10.cm]{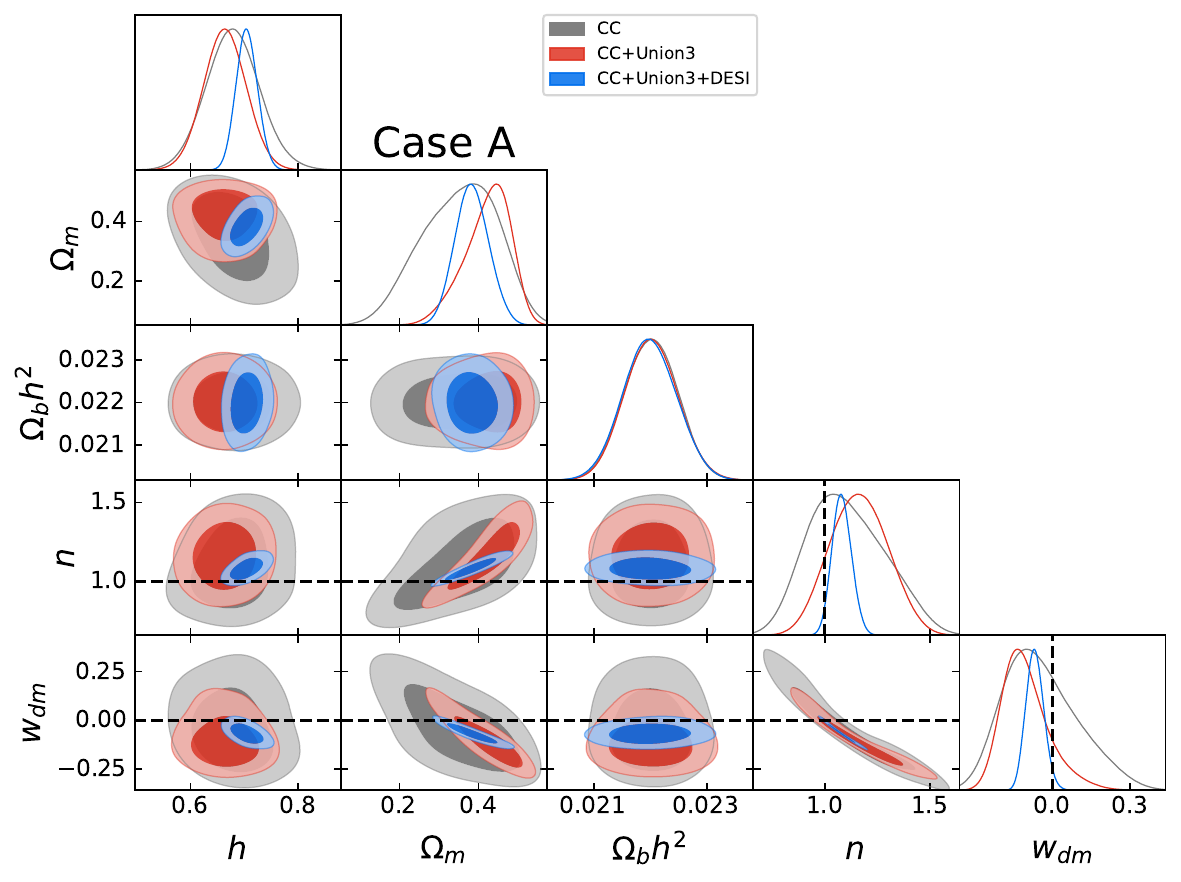}
    \includegraphics[width=9.cm]{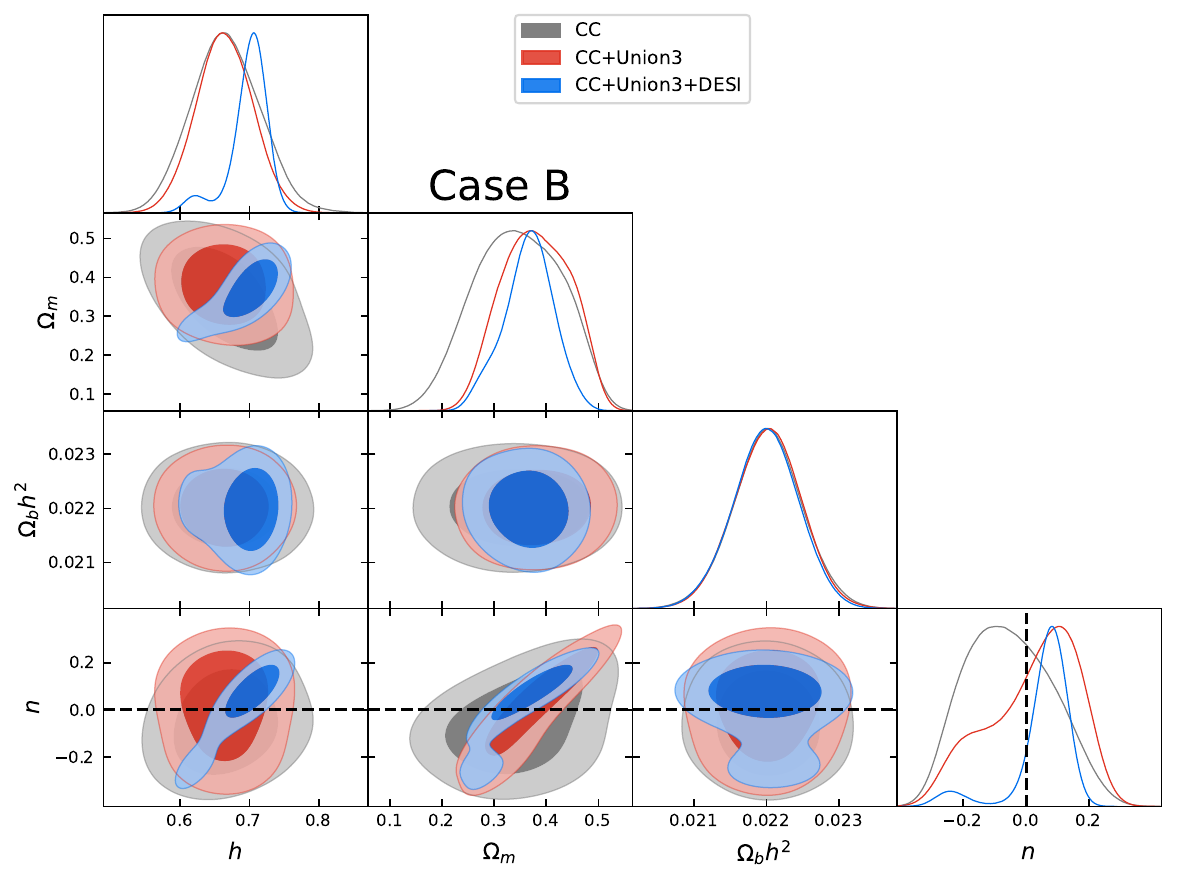}
    }
   \caption{Marginalized posterior distributions for the parameters of the Nonlinear Matter Coupling with Vacuum Term model (left) and the Nonlinear Matter Terms as Dynamical Dark Energy Mimickers model (right), at 1-$\sigma$ and 2-$\sigma$ confidence level. The vertical dashed lines represent the case of the standard $\Lambda$CDM model.}
   \label{fig:triangular}
\end{figure}

\begin{table}[htbp]
\centering
\begin{tabular}{lccc ccc}
\toprule
\multicolumn{1}{c}{} & \multicolumn{2}{ c }{CC} & \multicolumn{2}{ c }{CC+Union3} & \multicolumn{2}{ c }{CC+Union3+DESI}\\
\textbf{$\theta_i$}  & \textbf{Case A} & \textbf{Case B} & \textbf{Case A} & \textbf{Case B} & \textbf{Case A} & \textbf{Case B}  \\
\midrule
$\Omega_{m}$   
& $0.353 \pm 0.092$ & $0.346 \pm 0.082 $ & $0.423 \pm 0.056$ &  $0.380 \pm 0.063$& $0.383 \pm 0.041$ & $0.365 \pm 0.048$   \\
$h\times 100$            
& $67.86 \pm 4.88$  & $66.59 \pm 4.90$   & $66.39 \pm 3.88$   &  $66.42 \pm 4.05$ & $70.45 \pm 1.95$ & $69.82 \pm 2.92$ \\
$n$            
& $1.10 \pm 0.17$   & $-0.053 \pm 0.137$   & $1.16 \pm 0.13$     &  $0.023 \pm 0.143$ & $1.08 \pm 0.045$  & $0.054\pm 0.095$\\
\bottomrule
$\log(\mathcal{Z}_i)$ & $-7.22 \pm 0.12$ &  $-7.41 \pm 0.13$ & $-20.02 \pm 0.14$ & $ -20.41 \pm 0.14$& $-30.26 \pm 0.18$ & $-29.88 \pm 0.18$\\
$\chi^2_i$            &  5.94 & 6.03 & 29.28 & 28.99 &44.45 & 43.71 \\
\bottomrule

$\chi^2_{\Lambda}$ & \multicolumn{2}{ c }{6.11} & \multicolumn{2}{ c }{30.11} & \multicolumn{2}{ c }{47.18}\\
$\log(\mathcal{Z}_{\Lambda})$ & \multicolumn{2}{ c }{$-6.78 \pm 0.12$} & \multicolumn{2}{ c }{$-20.05 \pm 0.15$} & \multicolumn{2}{ c }{$-30.01 \pm 0.18$}\\
\bottomrule
$\Delta {\rm AIC}_{i, \Lambda}$  & 1.83 & 1.92 &  1.17 & 0.88 & -0.73 & -1.47 \\
$\Delta {\rm BIC}_{i, \Lambda}$  & 3.26 & 3.35 &  3.17 & 2.88 & 1.44 & 0.7 \\

$B_{i, \Lambda}$ & $-0.44\pm 0.17$ &  $-0.63\pm 0.17$ & $0.03\pm 0.20$ & $-0.36\pm 0.20$ & 
$-0.25\pm 0.25$ & $0.02\pm 0.25$\\
\bottomrule

\end{tabular}
\caption{Mean values along with  68\% confidence limits for cosmological parameters. 
Nonlinear Matter Coupling with Vacuum Term  model (Case A) and the Nonlinear Matter Terms as Dynamical Dark Energy Mimickers model (Case B).}
\label{tab:nested_summary}
\end{table}

In this section, we present and analyze the parameter inference results, for both cases A and B, and contrast them with the $\Lambda$CDM model, based on the data combinations described earlier. 
The parameter constraints are summarized in Figure \ref{fig:triangular}, which displays the combined 1D and
2D posterior distributions for the cosmological parameters across the models. 
The corresponding mean values are also listed in Table \ref{tab:nested_summary} with uncertainties reported at a confidence level of 68\%.

For both models, the Cosmic Chronometers, by themselves, provide very loose constraints on the cosmological parameters, as shown in the gray color of Fig. \ref{fig:triangular}. The constraints slightly improve when the SN-Union3 data set is incorporated (red), but remain statistically indistinguishable from $\Lambda$CDM. However, the addition of DESI-BAO considerably enhances the constraining power.  

For context, in the Nonlinear Matter Coupling with Vacuum Term (Case A), the associated equation of state is given by $w_{dm}=(1-n)/(2n-1)$, hence different values of $n=1$ represent possible deviations as the cold dark matter being described by dust. We output this parameter as a derived quantity and find a preference for slightly negative values, that is, for CC: $w_{dm}=-0.046\pm 0.139$, CC+Union3: $w_{dm}=-0.104\pm 0.088$ and CC+Union3+DESI: $w_{dm}=-0.067\pm 0.033$, which means the dark matter energy density decreases with cosmic expansion; posterior distributions for $w_{dm}$ are shown in the last row of the left panel in Fig. \ref{fig:triangular}. That is, in the $\Lambda$CDM context, the dark matter energy density behaves as $\rho_{dm}\propto(1+z)^3$, however, for this model, and using the last combination of data sets, it has a lower value in the exponent given by $2.798\pm 0.101$, indicating an earlier departure from the matter-dominated regime; in line with the results found in \cite{Escamilla:2023shf}. This result is corroborated by an improvement in the value of $\chi^2$ and slightly in favor by the AIC, but against the BIC and inconclusive through the Bayesian evidence, due to the inclusion of the additional $n$-parameter.  

\begin{figure}[h]
    \centering
     \makebox[12cm][c]{
    \includegraphics[trim = 0mm  0mm 0mm 0mm, clip, width=9.0cm, height=5.5cm]{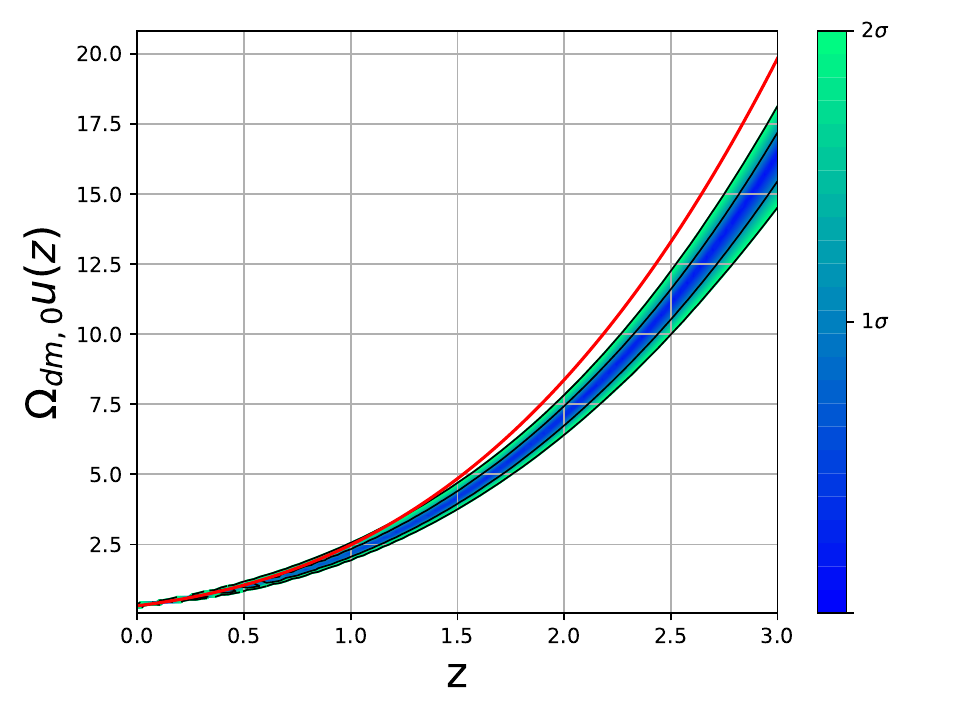}
    \includegraphics[trim = 0mm  0mm 0mm 0mm, clip, width=9.0cm, height=5.5cm]{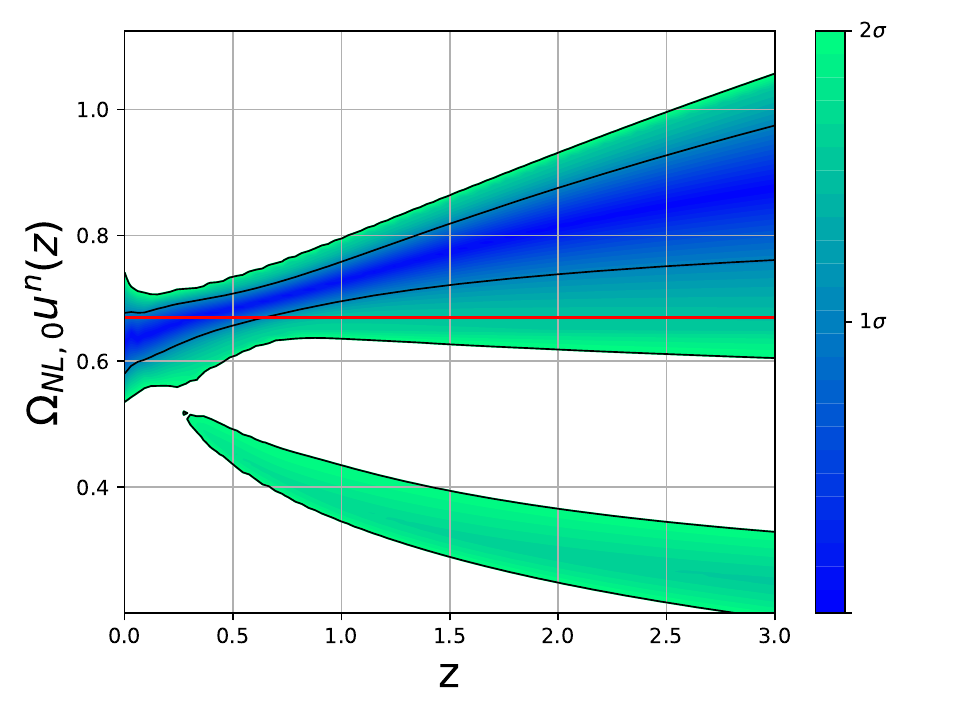}
    }
   \caption{Functional posterior of the, left, dark matter contribution $\Omega_{dm,0} u(z)$ and, right, the Non-Lineal term $\Omega_{NL,0} u^n(z)$, using the data set combination \textbf{CC+Union3+DESI}. The solid red lines indicate the $\Lambda$CDM model. The 68\% (1$\sigma$) and 95\% (2$\sigma$) confidence intervals are plotted as black lines.}
   \label{fig:posteriors}
\end{figure}

The results of the Nonlinear Matter Terms as Dynamical Dark Energy Mimicker model (Case B) can be summarized in Fig. \ref{fig:posteriors}, where we used the combination of the datasets CC+Union3+DESI. For comparison, the red lines indicate the $\Lambda$CDM model. In the left panel we display the dark matter contribution $\Omega_{dm,0} u(z)$ that again shows a decreasing behavior with the cosmic expansion. In the right panel the Non-Lineal term $\Omega_{NL,0} u^n(z)$ presents an increasing tendency through redshift, showing regions that deviate from the cosmological constant, for which the AIC and Bayesian evidence are slightly in favor of this model. An interesting result is a decreasing green contour arising from a second, minor, peak at negative values $n \sim -0.2$. Notice that in the marginalized posterior distributions (Fig. \ref{fig:triangular}), there exists a high correlation of the parameters $n$ and $h$, where small (negative) values of $n$ may match the constraints of Planck and hence contribute to ameliorate the Hubble tension, i.e., $n\sim -0.2$ may correspond to $h\sim 0.62$. A similar decrease in the energy density has been found using model-independent techniques that explain several datasets; see, for instance, \cite{Escamilla:2021uoj}.

\section{Conclusions}

We have investigated a class of $f(R,L_m)$ cosmologies in which nonlinear powers of the matter Lagrangian can reproduce late-time acceleration through matter--curvature interactions. Working with a Lagrangian density for dark matter as $L_m=\rho_m$ (and an uncoupled Maxwell radiation sector), we analyzed two representative realizations and combined a global phase-space study with background-only Bayesian constraints. For Case A, $f(R,L_m)=R/2+\beta\rho_m^{n}+\gamma$, the bounded dynamical system shows that the standard cosmic sequence radiation$\to$matter$\to$deSitter is recovered only for $n\gtrsim 4/5$ within the physically admissible sector.
In this model, accelerated expansion is not generated as a genuinely emergent outcome of the nonlinear matter term alone: the late-time de~Sitter point is primarily controlled by the explicit vacuum contribution $\gamma$, while the exponent $n$ mainly regulates the intermediate-era dilution law and the timing of the approach to the attractor. Consistently, the background fit with CC+Union3+DESI favors values close to the $\Lambda$CDM limit, $n=1.08\pm0.05$ (68\% CL), with model-comparison statistics indicating at most a mild preference depending on the criterion (AIC vs.\ BIC/evidence).

For Case B, $f(R,L_m)=R/2+\beta\rho_m+\gamma\rho_m^{n}$, the dynamical-systems analysis reveals a qualitatively different mechanism. A complete radiation$\to$matter$\to$acceleration sequence inside the bounded physical simplex arises only in the window $0<n<1/2$, where the system possesses a \emph{scaling} de~Sitter future attractor (with $q=-1$ and $\omega_{\rm eff}=-1$) at nonzero $(\Omega_m,\Omega_{NL})$. In this regime, acceleration occurs without introducing an explicit vacuum term in the action, and the nonlinear density contribution behaves as an effective, density-dependent coupled sector that can dominate at late times. The background data constrain $n=0.05\pm0.10$ (68\% CL) for CC+Union3+DESI, i.e.\ values consistent with the accelerating branch while remaining statistically compatible with $\Lambda$CDM kinematics at the background level. This is consistent with the fact that background probes are largely insensitive to the detailed composition of the effective sector, underscoring the importance of complementary tests.
We have also recorded minimal consistency requirements relevant for viability. In the present realizations $f_R=1/2>0$, so the tensor sector is free of ghost instabilities and the effective gravitational coupling does not change sign; moreover, the absence of derivative tensor couplings implies luminal propagation, $c_T^2=1>0$, excluding tensor gradient instabilities.

We further require $f_{L_m}>0$ over the relevant density range to avoid pathological sign flips in the effective matter--geometry coupling.
These conditions are necessary but not sufficient: a definitive assessment requires the full linear perturbation theory, including scalar no-ghost/no-gradient criteria, the rest-frame sound speed, and the effective modified-gravity functions $\mu(k,a)$, $\Sigma(k,a)$ and $\eta(k,a)$ to be confronted with growth ($f\sigma_8$) and lensing data.
Therefore, Case A provides a controlled extension closely connected to $\Lambda$CDM in which acceleration is vacuum-driven, whereas Case B singles out a genuinely distinctive branch where late-time acceleration emerges dynamically from nonlinear matter--curvature coupling. The observational results presented here motivate a dedicated perturbation-level study to determine whether the accelerating window in Case B remains viable when structure formation and lensing constraints are included.

\section*{Acknowledgements}

We thank the support of the FORDECYT-PRONACES-CONACYT project CF-MG-2558591 grant. R. G-S acknowledges support from the Secretaría de Investigación y Posgrado of the Instituto Politécnico Nacional for funding this work (projects SIP20240638 and SIP20250043). J.A.V. acknowledges support from FOSEC SEP-CONACYT Ciencia Básica A1-S-21925,  UNAM-DGAPA-PAPIIT IN117723, IN110325, and Cátedra de Investigación Marcos Moshinsky. The authors thank Miguel A Zapata for useful discussions on the observational analysis.

\bibliographystyle{unsrtnat}
\bibliography{f_RLm_bib}

\end{document}